\documentclass[twocolumn]{aastex7}

\usepackage{booktabs}
\usepackage{amsmath}
\usepackage{booktabs}
\usepackage{caption}
\usepackage{threeparttable}
\usepackage[table]{xcolor}
\definecolor{wdblue}{RGB}{230,238,250}
\definecolor{msred}{RGB}{255,235,235}
\definecolor{systemgray}{RGB}{235,235,235}
\usepackage[table]{xcolor}
\usepackage{threeparttable}
\usepackage{caption}
\usepackage{booktabs} 
\usepackage{amsmath}
\usepackage{booktabs}
\usepackage{threeparttable}
\usepackage{graphbox}
\usepackage[utf8]{inputenc}

\defcitealias{Grondin2024}{G24}

\shorttitle{Linking Pre/Post-CE Binary Properties with Star Clusters}
\shortauthors{Grondin et al.}

\begin{document}

\title{A Framework for Linking Pre- and Post-Common Envelope Binary Properties with Star Clusters: \\ The First Demonstration with a Massive White Dwarf+M Dwarf Binary in Alessi 12}

\author[0000-0002-0444-8502]{Steffani M. Grondin}
\affiliation{Department of Astronomy \& Astrophysics, University of California San Diego, La Jolla, CA 92093, USA}
\affiliation{David A. Dunlap Department of Astronomy \& Astrophysics, University of Toronto, 50 St. George St., Toronto, ON M5S 3H4, Canada}
\email{sgrondin@ucsd.edu}

\author[0000-0001-7081-0082]{Maria R. Drout}
\affiliation{David A. Dunlap Department of Astronomy \& Astrophysics, University of Toronto, 50 St. George St., Toronto, ON M5S 3H4, Canada}
\email{maria.drout@utoronto.ca}

\author[0000-0002-5608-4683]{Jason Nordhaus}
\affiliation{National Technical Institute for the Deaf, Rochester Institute of Technology, 1 Lomb Memorial Dr., Rochester, NY 14623, USA}
\affiliation{Center for Computational Relativity and Gravitation, Rochester Institute of Technology, \\ 1 Lomb Memorial Dr., Rochester, NY 14623, USA}
\email{jtnsma@rit.edu}

\author[0000-0002-0638-8822]{Philip S. Muirhead}
\affiliation{Department of Astronomy \& The Institute for Astrophysical Research, Boston University, \\
725 Commonwealth Ave., Boston, MA 02215, USA}
\email{philipm@bu.edu}

\author[0000-0002-7647-3555]{Bailey Filer}
\affiliation{Center for Computational Relativity and Gravitation, Rochester Institute of Technology, \\ 1 Lomb Memorial Dr., Rochester, NY 14623, USA}
\affiliation{School of Physics and Astronomy, Rochester Institute of Technology, 1 Lomb Memorial Dr., Rochester, NY 14623, USA}
\email{bnf4525@rit.edu}

\author[0000-0002-5522-0217]{Alexander Laroche}
\affiliation{David A. Dunlap Department of Astronomy \& Astrophysics, University of Toronto, 50 St. George St., Toronto, ON M5S 3H4, Canada}
\affiliation{Dunlap Institute for Astronomy \& Astrophysics, University of Toronto, 50 St George Street, Toronto, ON M5S 3H4, Canada}
\email{alex.laroche@mail.utoronto.ca}

\author[0000-0003-3613-0854]{Jeremy J. Webb}
\affiliation{Department of Science, Technology, and Society, Division of Natural Sciences, York University, \\ 218 Bethune College, Toronto, ON, M3J 1P3, Canada}
\email{webbjj@yorku.ca}

\author[0000-0002-4421-4962]{Floor S. Broekgaarden}
\affiliation{Department of Astronomy \& Astrophysics, University of California San Diego, La Jolla, CA 92093, USA}
\email{fbroekgaarden@ucsd.edu}

\author[0000-0002-7706-5668]{Ryan Chornock}
\affiliation{Department of Astronomy, University of California, Berkeley, CA 94720-3411, USA}
\affiliation{Berkeley Center for Multi-messenger Research on Astrophysical Transients and Outreach (Multi-RAPTOR), \\ University of California, Berkeley, CA 94720-3411, USA}
\email{chornock@berkeley.edu}

\author[0000-0002-4086-3180]{Kyle Kremer}
\affiliation{Department of Astronomy \& Astrophysics, University of California San Diego, La Jolla, CA 92093, USA}
\email{kykremer@ucsd.edu}

\author[0000-0002-2249-0595]{Natalie LeBaron}
\affiliation{Department of Astronomy, University of California, Berkeley, CA 94720-3411, USA}
\affiliation{Berkeley Center for Multi-messenger Research on Astrophysical Transients and Outreach (Multi-RAPTOR), \\ University of California, Berkeley, CA 94720-3411, USA}
\email{nlebaron@berkeley.edu}

\author[0000-0003-4768-7586]{Raffaella Margutti}
\affiliation{Department of Astronomy, University of California, Berkeley, CA 94720-3411, USA}
\affiliation{Berkeley Center for Multi-messenger Research on Astrophysical Transients and Outreach (Multi-RAPTOR), \\ University of California, Berkeley, CA 94720-3411, USA}
\email{rmargutti@berkeley.edu}

\author[0009-0001-9577-9912]{Nikki Noughani}
\affiliation{David A. Dunlap Department of Astronomy \& Astrophysics, University of Toronto, 50 St. George St., Toronto, ON M5S 3H4, Canada}
\affiliation{Dunlap Institute for Astronomy \& Astrophysics, University of Toronto, 50 St George Street, Toronto, ON M5S 3H4, Canada}
\email{n.noughani@utoronto.ca}

\author[0000-0001-8023-4912]{Huei Sears}
\affiliation{Department of Physics and Astronomy, Rutgers, the State University of New Jersey, \\ 136 Frelinghuysen Road, Piscataway, NJ 08854-8019, USA}
\email{hs1349@physics.rutgers.edu}

\author[0000-0001-9873-0121]{Pier-Emmanuel Tremblay}
\affiliation{Department of Physics, University of Warwick, Coventry CV4 7AL, UK}
\email{P.Tremblay@warwick.ac.uk}

\correspondingauthor{Steffani M. Grondin}
\email{sgrondin@ucsd.edu}

\begin{abstract}

Common envelope (CE) evolution is a critical phase in the lives of binary stars, producing close binaries that are progenitors of type Ia supernovae and gravitational wave sources. Despite its importance, CE evolution remains poorly understood, largely due to the 
scarcity of systems with constrained pre- and post-CE properties. Here, we present a star cluster-based framework for reconstructing the evolutionary histories of white dwarf+main-sequence (WD+MS) post-CE binaries, where cluster membership can provide an independent age constraint and/or rule out a merger origin for the WD. We demonstrate this method with Alessi12-PCE, the first such binary in an open cluster with precisely determined pre- and post-CE properties. We classify the companion as an M4V and measure a WD mass of $1.06 \pm 0.02 M_{\odot}$, making it the most massive WD+MS binary associated with a cluster. A 6.99-hour periodicity detected in a light curve is confirmed as the binary orbital period via radial velocity monitoring. Combined with the WD mass, WD cooling age, and Alessi 12 cluster age, stellar evolution models imply a $5.40 \pm 0.10 M_{\odot}$ WD progenitor that entered a CE on the asymptotic giant branch (AGB). CE evolution models where convection is the dominant physical mechanism that sets $\alpha_{\text{CE}}$ reproduce the observed orbital separation in exactly two scenarios: either a mid-AGB interaction with $\alpha_{\text{CE}}\approx0.99$, or a late-AGB interaction with $\alpha_{\text{CE}}\approx0.05$. Applicable to other post-CE binaries in star clusters, our new framework enables empirical constraints on CE physics inaccessible from field binaries alone. 

\end{abstract}


\vspace{-20pt}

\section{Introduction} \label{sec:intro}

Common envelope (CE) evolution is one of the least understood, yet most important phases in the lives of binary stars. 
Despite lasting only on the order of hundreds of days \cite[][]{WilsonNordhaus2019}, the CE phase is theorized to be a crucial mechanism for producing close binaries from initially wide systems \citep{Ivanova2013,  Ivanova:CEbook}. It therefore plays a key role in the formation of cataclysmic variables \citep[e.g.,][]{Paczynski1976, Warner1995}, X-ray binaries \citep[e.g.,][]{TaurisvandenHeuvel2006, TaurisvandenHeuvel2023}, gravitational wave sources \citep[e.g.,][]{Belczynski2017, Broekgaarden2026}, Type Ia supernovae \citep[e.g.,][]{Webbink1984,
LiuRopkeHan2023}, and gamma-ray bursts \citep[e.g.,][]{FryerWoosleyHartmann1999, TaurisvandenHeuvel2023}.
A CE phase \citep{Paczynski1976} typically occurs when a giant star overfills its Roche lobe and unstable mass transfer causes its companion to become engulfed within a shared envelope. Drag forces inside the envelope remove angular momentum and orbital energy from the binary, driving a rapid inspiral. If sufficient energy is transferred to the envelope, it is ejected, leaving a compact post-CE binary; otherwise, the stellar cores merge into a single star \citep[see work by e.g.,][]{2025arXiv250200111S}. 

Although many binaries are expected to undergo CE evolution, its detailed physics and eventual outcomes remain uncertain. This reflects both the intrinsic complexity of CE evolution and the computational challenge of modeling it: the interaction spans many physical processes, spatial scales, and evolutionary timescales, making fully self-consistent 3D simulations prohibitively expensive \citep{Ropke2023}. Consequently, both rapid  (e.g., \texttt{BSE}, \texttt{COSMIC}, \texttt{COMPAS}; \citealt{2002MNRAS.329..897H, 2020ApJ...898...71B, 2022ApJS..258...34R}) and more detailed stellar-structure (e.g., \texttt{BPASS}, \texttt{POSYDON}; \citealt{2017PASA...34...58E, 2023ApJS..264...45F, 2025ApJS..281....3A}) binary population synthesis codes rely on heuristic approaches, such as the simplified CE efficiency parameter, $\alpha_{\text{CE}}$, and the $\alpha \lambda$ prescription to approximate CE evolution outcomes \citep[e.g.,][]{Paczynski1976, Webbink1984, Ivanova:CEbook}. 
The $\alpha_{\text{CE}}$ parameter quantifies the efficiency with which the change in orbital energy of the binary ($\Delta E_{\text{orb}}$) is converted into the energy required to unbind the donor's envelope (whose binding energy is $E_{\text{bind}}$).

While $\alpha_{\text{CE}}$ is currently the most widely used parameter to describe the outcomes of CE evolution, it is a highly simplified, energy-based formalism that reduces a complex, large-scale process to a single number.
Alternative analytic prescriptions have been proposed to address some of these challenges, including the $\gamma$-formalism which considers angular momentum rather than energy budgets \citep{Nelemans2000}, and `SCATTER', which models the behavior of different parts of the envelope during a CE event \citep{DiStefano2023}. Unfortunately, none of these formalisms capture the true complexity of the CE phase, where the methods remain 1D, parameterized approximations rather than physical models of the interaction itself. As such, these simplified parameterizations often struggle to reproduce observed post-CE binary populations, likely due to missing physics, such as convection \citep{WilsonNordhaus2019, 2020MNRAS.497.1895W, 2022MNRAS.516.2189W, Reichardt2020, Lau2022}. This uncertainty is highlighted by recent discoveries from \textit{Gaia}, for example, which have revealed dozens of white dwarf (WD)+MS post-CE binaries with unexpectedly long orbital periods \citep[$\sim$100–1000 days;][]{2024PASP..136h4202Y, 2024MNRAS.52711719Y, 2025arXiv250514786Y}.  

This limitation has motivated 3D hydrodynamic simulations that attempt to directly model, rather than parameterize, the CE interaction \citep{Passy2012, Ohlmann2016, LawSmith2020, Lau2022}. However, as stated above, while these simulations offer a more physically-motivated picture of the interaction, their computational expense limits them to a small number of donor masses and orbital configurations, making them  impractical for the large-scale population synthesis studies that rely on the parameterized prescriptions above.

One of the largest challenges in attempting to improve our understanding of CE evolution is that there are very few observed binary systems in which both the post-CE and pre-CE parameters (e.g., initial and final masses and orbital separations) are precisely constrained. Most post-CE binaries are field stars without precise age determinations. Uncertainties in the WD cooling tracks prevent an accurate determination of the WD progenitor's initial mass and cannot rule out a previous WD merger. In other words, for post-CE field systems, we cannot distinguish whether the WD originated from: (i) the core of a more massive star earlier in its evolution, (ii) the core of a lower mass star later in its evolution, or (iii) a merger within a multiple-star system taking place well after initial formation. An independent stellar age constraint is needed to accurately determine the pre-CE binary parameters and trace the full evolution of the system. However, if both the pre- and post-CE parameters of systems are known, numerical and hydrodynamical simulations could be anchored with initial conditions and final outcomes to reproduce, and binary population synthesis efforts could even potentially bypass heuristic prescriptions altogether by directly mapping initial configurations to final separations. 

Open star clusters provide one example of a stellar population where constraining the pre-CE masses and initial binary separation may be feasible, as these systems typically have well-constrained ages via isochrone fitting \citep[e.g.,][]{2019A&A...623A.108B, CG2020,2021ApJS..257...46G} or gyrochronology \citep[e.g.,][]{Bouma...NGC2516...2021AJ....162..197B,Fritzewski...NGC2281...2023A&A...674A.152F, 2025ApJ...986...59V}. A precisely measured cluster age, combined with the WD's cooling age, constrains the lifetime of the WD progenitor, and thus its initial mass, via stellar evolution models. Moreover, for clusters with a turn-off mass exceeding $1.5\,\text{M}_\odot$, the CE most likely occurred during the asymptotic giant branch (AGB) phase of the primary star. This is because, for stars of these masses, the AGB radius is significantly larger than at any earlier evolutionary phase, making Roche lobe overflow (and thus a CE) more likely to occur at this stage. Since the WD mass remains within a few percent of its final value during the AGB phase \citep{2015MNRAS.446.2599D, 2020NatAs...4.1102M,2022Univ....8..243M}, the post-CE WD mass is a good approximation for its mass at CE onset, allowing the progenitor's initial mass to be accurately determined from the WD's current post-CE mass, cooling age, and cluster age. This is analogous to how the initial-final mass relation (IFMR) is determined for isolated WDs in star clusters \citep[e.g.,][]{Cummings2018, 2021ApJ...912..165R, 2022ApJ...926L..24M, 2026ApJ...996...69M}. Furthermore, an observed WD mass or temperature inconsistent with the age of the star cluster may indicate that the WD was produced by a merger rather than regular evolution, offering a diagnostic for distinguishing between these two outcomes.

Despite the past and ongoing efforts that have identified thousands of WD+MS binaries in our Milky Way \citep[][]{2007MNRAS.382.1377R, RM2010, Ren2014, 2016MNRAS.463.2125P, 2017MNRAS.470.1442C, RM2021, Grondin2024, 2024arXiv240407388J, Nayak2024, 2024MNRAS.529.3729S, Li2025, 2026PASP..138c4202S, 2026ApJS..284...38Y}, only three detached WD+MS systems have ever been associated with a star cluster, and only two are confirmed post-CE binaries\footnote{\cite{2019ApJ...880...75R} discovered a massive ($1.06 M_{\odot}$) WD+M dwarf binary in the open cluster NGC 2422, however the system does not yet have a measured orbital period, making it unclear whether it underwent a CE phase and can therefore be classified as a post-CE binary.}, likely due to the historical lack of astrometric data prior to the \textit{Gaia} space mission \citep{2016A&A...595A...2G}. These post-CE binaries are V471 Tau \citep{1987AJ.....94..996S} and HZ9 \citep{1971ApJ...166L..81Y}, which are both located in the Hyades star cluster. Although V471 Tau is a confirmed post-CE binary, it is unsuitable for the method described above: analysis by \cite{2022AJ....163...34M}, building on \cite{2001ApJ...563..971O}, suggests the WD is too massive and hot for its age in the Hyades. Instead, a merger within a triple system is favored, with the resulting product later entering a CE with the tertiary star, making it difficult to determine the precise pre-CE conditions.  Although new constraints on HZ9 make it a promising system for studying CE evolution with low-mass WDs \citep{Muirhead2024}, it remains just a single example. Discovering and characterizing additional post-CE binaries across different cluster environments, masses, and separations is crucial for providing robust observational constraints on CE evolution.

\begin{figure*}[!ht!]
    \centering
    \includegraphics[width=1\linewidth]{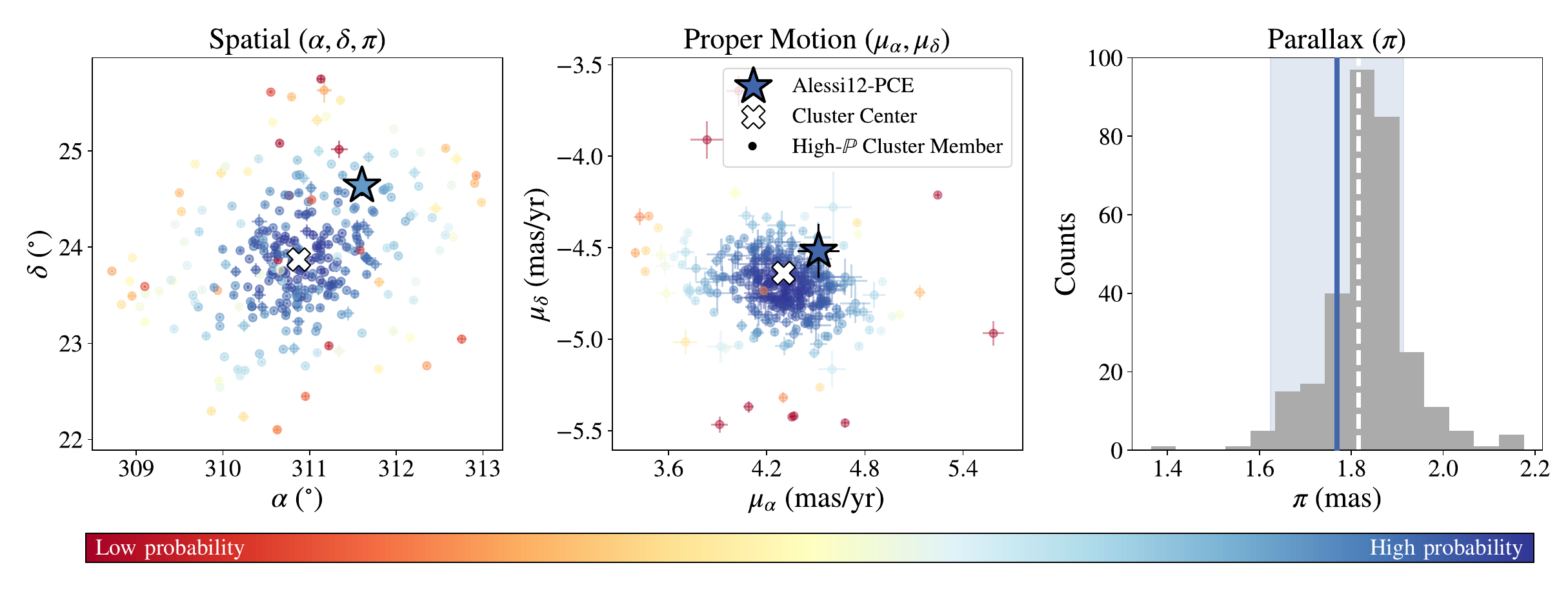}
    \caption{A $\chi^{2}$ cluster membership analysis for Alessi12-PCE. \textit{Left:} A 3D spatial analysis of Alessi12-PCE (star) relative to the probable cluster members ($P > 0.5$) of Alessi 12 from \cite{CG2020} (circles, colored by their $\chi^{2}$ value), with the cluster center marked by a white cross. Each source is colored by its spatial $\chi^{2}$ value (color bar), computed from a combination of right ascension ($\alpha$), declination ($\delta$), and parallax ($\pi$), following the methodology of \citetalias{Grondin2024}: lower $\chi^{2}$ (bluer) indicates closer consistency with the cluster.  
    \textit{Middle:} The same,  but for a 2D proper motion ($\mu_\alpha$, $\mu_\delta$)  cluster membership analysis. 
    \textit{Right:} A parallax ($\pi$) distribution of the same \cite{CG2020} cluster members (gray histogram), with Alessi12-PCE's parallax (blue line) and its uncertainty (light blue shading) along with the cluster parallax (white dashed line) indicated. The spatial location and kinematics of Alessi12-PCE are highly consistent with those of high-probability members of the Alessi 12 star cluster.}
    \label{fig:alessi12kin}
\end{figure*}

In this work, we present a new framework for reconstructing the evolutionary histories of WD+MS post-CE binaries using star clusters. We detail this framework in the context of Alessi12-PCE -- the first post-CE binary in a star cluster for which both the pre-CE and post-CE system parameters can be precisely constrained. Alessi12-PCE is a newly identified WD+MS binary in the open cluster Alessi 12, discovered in the \cite{Grondin2024} (hereafter G24) catalog of WD+MS candidate binaries in star clusters. As only the third detached WD+MS post-CE binary ever associated with a star cluster, it provides a unique observational benchmark for constraining and testing models of CE evolution.  
In Section \ref{sec:discovery}, we summarize the discovery of Alessi12-PCE. In Section \ref{sec:obs}, we detail our multi-wavelength observations and data reductions. In Section \ref{sec:analysis}, we describe our spectral fitting routine to determine the M dwarf and WD properties. In Section \ref{sec:orbits}, we present the orbital parameters of Alessi12-PCE. In Section \ref{sec:masses}, we determine the post-CE masses of both the M dwarf and WD, while in Section \ref{sec:preCE}, we infer the pre-CE progenitor masses. In Section \ref{sec:discussion}, we examine the CE physics that produced Alessi12-PCE, where we compute a range of $\alpha_{\text{CE}}$ values for this system, and discuss Alessi12-PCE in the context of physically motivated convective CE models by \cite{WilsonNordhaus2019}. Finally, we discuss alternative binary formation pathways in Section \ref{sec:alternate} and conclude in Section \ref{sec:conclusions}.

\section{Discovery} \label{sec:discovery}

Alessi12-PCE was recently discovered as `Alessi12-c1' in the \citetalias{Grondin2024} catalog of candidate WD+MS binaries in star clusters. In this work, \citetalias{Grondin2024} implemented a supervised machine learning method \citep[a support vector machine;][]{Cortes1995} to search hundreds of open clusters for systems that had similar photometric colors to a sample of confirmed WD+MS field binaries. Specifically, they used the OC catalog of \cite{CG2020}, the SDSS WD+MS sample from \cite{RM2010}, and a combination of \textit{Gaia} \citep{2023A&A...674A...1G}, Pan-STARRS1 \citep{2016arXiv161205560C}, and 2MASS \citep{2006AJ....131.1163S} photometry. A total of 52 high-probability WD+MS candidate binaries in 38 open clusters were identified. Of these, Alessi12-PCE was the bluest source in the sample ($G_{\rm BP} - G_{\rm RP}=1.093$ mag) and had the highest probability ($P=0.999$) of being a WD+MS binary. This system was identified as a member of the Alessi 12 star cluster by comparing its position, proper motion, and parallax to confirmed Alessi 12 members from \cite{CG2020}, as we show in Figure \ref{fig:alessi12kin}. Alessi12-PCE has no other classifications in the SIMBAD astronomical database \citep{2000A&AS..143....9W}. The coordinates, parallax, proper motion, and archival photometry of Alessi12-PCE are listed in Table~\ref{tab:sysprops}. The broadband photometry is discussed further in Section~\ref{sec:broadband}.

\citetalias{Grondin2024} confirmed Alessi12-PCE to be a WD+MS binary through a spectrum obtained with the Kast spectrograph mounted on the Shane telescope at the Lick Observatory \citepalias[Figure 10 in][]{Grondin2024}. This spectrum revealed a very strong blue continuum between $\sim 3500-5000$~\AA~with broad Balmer absorption lines, indicating the presence of a WD in the system. However, at redder wavelengths ($>6000$ \AA) the spectrum shows a red continuum with molecular absorption bands which establish the presence of an M-type MS star. A light curve obtained from the Zwicky Transient Facility (ZTF) database also revealed variability with an amplitude of $\sim 0.2$ mag (r-band) and a clear $\sim 0.29$ day periodicity, potentially indicating that the system is a short-period binary.

\begin{table}[!th]
\centering
\caption{\textit{Gaia} DR3 astrometry and archival photometry of Alessi12-PCE.}
\label{tab:sysprops}
\renewcommand{\arraystretch}{1.15}
\begin{tabular}{llc}
\hline
Parameter & Value & Source \\
\hline
\multicolumn{3}{c}{\textit{Gaia} DR3 Astrometry} \\
\hline
\textit{Gaia} DR3 ID  & 1843297090388761600 & \textit{Gaia} DR3 \\
R.A. (J2016)          & \texttt{20:46:25.0} & \textit{Gaia} DR3 \\
Dec. (J2016)          & \texttt{+24:38:20.2} & \textit{Gaia} DR3 \\
Parallax $\varpi$     & $1.768 \pm 0.144$ mas & \textit{Gaia} DR3 \\
$\mu_{\alpha}$      & $4.516 \pm 0.129$ mas\,yr$^{-1}$ & \textit{Gaia} DR3 \\
$\mu_{\delta}$        & $-4.517 \pm 0.147$ mas\,yr$^{-1}$ & \textit{Gaia} DR3 \\
RUWE  & $0.950$               & \textit{Gaia} DR3 \\ \hline
\multicolumn{3}{c}{Archival Photometry [AB mag]} \\
\hline
$G$                       & $18.399 \pm 0.003$ & \textit{Gaia} DR3 \\
$G_{\rm BP} - G_{\rm RP}$ & $1.093 \pm 0.029$  & \textit{Gaia} DR3 \\
$g$             & $18.837 \pm 0.007$  &Pan-STARRS1 \\
$r$             & $18.724 \pm 0.008$  & Pan-STARRS1 \\
$i$             & $18.169 \pm 0.008$  & Pan-STARRS1 \\
$z$             & $17.635 \pm 0.010$ & Pan-STARRS1 \\
$y$             & $17.399 \pm 0.006$  & Pan-STARRS1 \\
$J$                       & $16.217 \pm 0.120$  & 2MASS \\
$H$                       & $15.756 \pm 0.150$  & 2MASS \\
$K_s$                     & $15.235 \pm 0.157$  & 2MASS \\
\hline
\end{tabular}
\end{table}

\begin{figure*}
    \centering
    \includegraphics[width=1\linewidth]{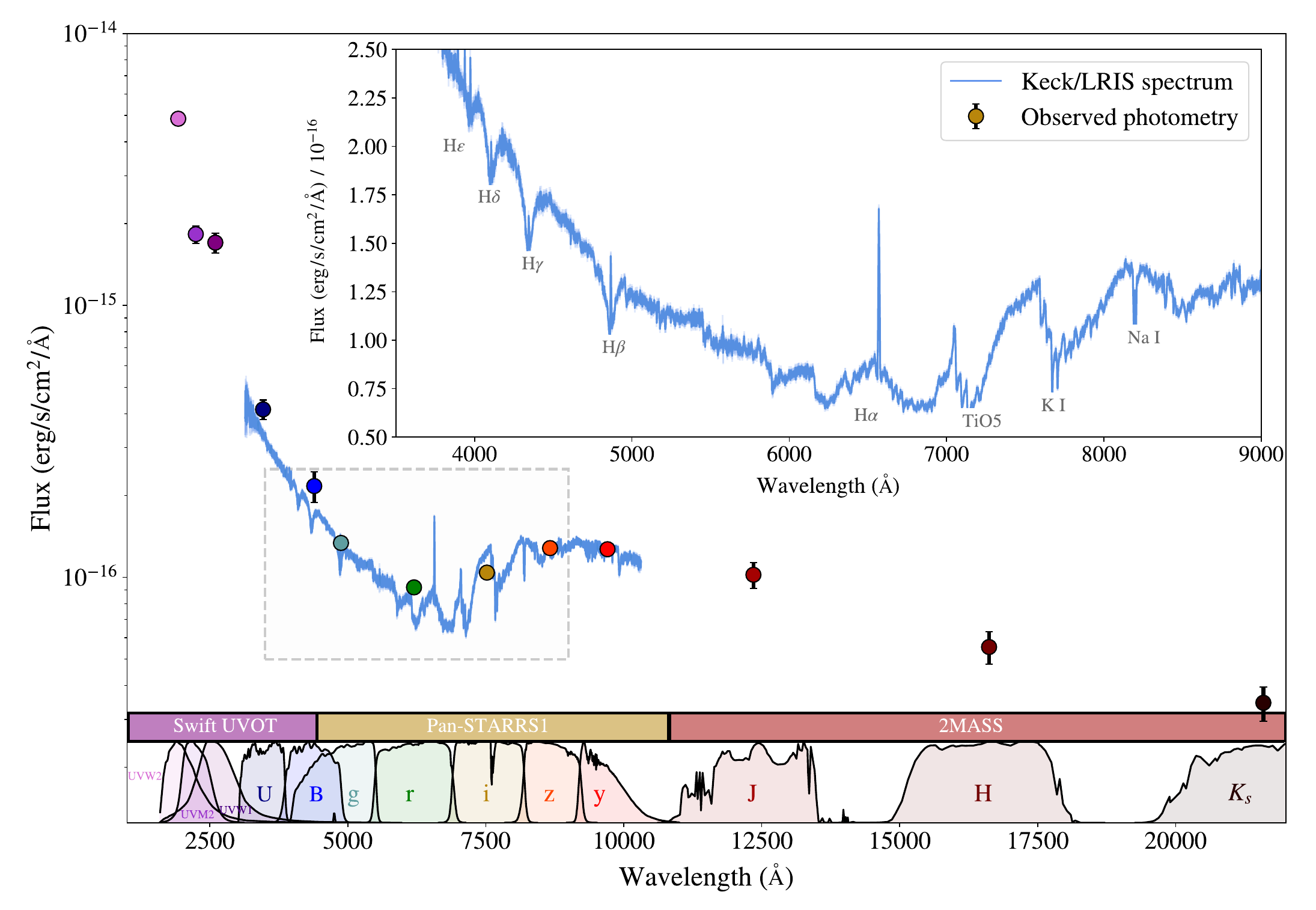}
    \caption{Multi-wavelength spectral energy distribution of Alessi12-PCE, combining ultraviolet photometry from \textit{Swift} UVOT (U, B, UVW1, UVM2, UVW2), optical photometry from Pan-STARRS1 ($g, r, i, z, y$), and near-infrared photometry from 2MASS ($J, H,$ $K_{s}$). Each photometric band is indicated accordingly. The SED reveals the composite nature of the system: a hot WD dominates the ultraviolet and blue optical flux, while an M dwarf companion dominates at red optical and near-infrared wavelengths. A Keck/LRIS spectrum of Alessi12-PCE is also shown, where a zoomed in region (dashed box) highlights the major WD and MS spectral features. Broad Balmer absorption lines (H$\alpha$, H$\beta$, H$\gamma$, H$\delta$ and H$\epsilon$) from the WD and molecular bands (TiO5, KI, and NaI) from the MS star are labeled. Neither the broadband photometry nor the spectrum has been corrected for interstellar reddening.}
    \label{fig:observedWDspectrum}
\end{figure*}

\section{Observations and Data Reduction}
\label{sec:obs}

A complete characterization of Alessi12-PCE requires as much observational coverage as possible: spectroscopy to resolve features from both the WD and MS star, broadband photometry to confirm the WD's ultraviolet (UV) continuum, and light curves to probe variability. In the subsequent sections, we describe each of these datasets used to analyze Alessi12-PCE.

\subsection{Spectroscopy}

Between 2023 and 2024, we obtained a set of high signal-to-noise (SNR) and/or multi-epoch spectroscopy of Alessi12-PCE on a number of instruments. Below, we describe the data acquisition and reduction, where Table \ref{tab:obssummary} presents a summary of these observations.

\subsubsection{Keck/LRIS}

We observed Alessi12-PCE on November 17, 2023 with the Low-Resolution Imaging Spectrometer \citep[LRIS;][]{lrisref} on the 10-meter Keck I telescope. We used the 600/4000 grism, a D560 dichroic beam splitter, and the 400/8500 grating to cover the wavelength range of 3140--10300~\AA. The exposure times were 1$\times$900s and 2$\times$425s on the blue and red sides, respectively. The spectroscopic reduction followed standard procedures (e.g., \citealt{silverman12}). Relative flux calibration was performed using an observation of EG131 \citep{bessell99} obtained on the same night, although variable clouds made the absolute fluxes unreliable without scaling to the photometry of the source. We plot this Keck/LRIS spectrum in Figure \ref{fig:observedWDspectrum}. A series of Balmer
lines (H$\beta$, H$\gamma$, H$\delta$, and H$\epsilon$) in absorption are visible, indicating the presence of a WD. We also observe emission lines in the cores of the Balmer absorption lines, indicating chromospheric emission. A series of molecular bands (e.g., TiO5, K I, and Na I) at redder wavelengths are seen, indicating the presence of an M-type MS star in the system.

\subsubsection{Gemini/GMOS}\label{sec:geminispec}

We obtained spectra of the Alessi12-PCE system with the Gemini Multi-Object Spectrograph (GMOS; \citealt{GMOS2004}) on the 8.1-meter Gemini-North telescope on nine different epochs between April 30 and July 12, 2024. All observations were obtained with the B480 grating, which offers wavelength coverage of $\sim$3700-7300 \AA\ at a resolution of $R \sim 1500$. Each epoch was typically divided into four observations obtained sequentially (two each at two central wavelengths: 520 and 530 nm) with exposure times between 690 and 705 seconds. 
  
All GMOS data were reduced using standard \texttt{gemini} and \texttt{gmos} tasks within   \texttt{pyraf} \citep{pyraf}. This process includes bias/overscan correction, flat-fielding, wavelength calibration, spectral extraction, and flux calibration. Flux calibration was performed using observations of the standard stars G191B2B and Feige66 obtained with the same set-up as our observations. After flux calibration, we stacked the 2--4 individual exposures for each epoch (as listed in Table~\ref{tab:obssummary}). In Section~\ref{sec:orbits}, below, we use both these nine stacked spectra and all 33 individual exposures when measuring the radial velocity (RV) shifts of the WD and M dwarf in the Alessi12-PCE system. 

In Figure~\ref{fig:gemini-allspec}, we plot the nine stacked Gemini/GMOS spectra. In general, similar features are observed as in the higher SNR Keck/LRIS spectrum shown in Figure~\ref{fig:observedWDspectrum}. However, we note that both the strength and number of narrow emission line features significantly increases during the observations obtained on May 30, June 1, and July 9, 2024 (compared to both the Keck spectrum obtained in November 2023 and the other Gemini observations obtained in the month(s) prior and after these dates). In particular, many additional emission lines  aside from the Balmer lines visible in Figure~\ref{fig:observedWDspectrum} are also observed.

\subsubsection{Shane/Kast}

We observed Alessi12-PCE with the Kast Double Spectrograph \citep{millerstone93} on the 3-meter Shane telescope at the Lick Observatory on three nights between October 24 and December 5, 2023. Our observations were taken with a 2$\arcsec$ slit, a d57
dichroic, the 600/4310 grism on the blue channel, and a 600/7500 grating on the
red channel, providing a resolution of $\approx 5$ \AA~over the range of $3520–8750$  \AA. Individual exposure times were 1830s on the blue channel and 900s on the red channel, as summarized in Table~\ref{tab:obssummary}. Data reduction (including bias/flat field correction, background subtraction, spectral extraction, and wavelength calibration) were performed using standard routines within \texttt{pyraf}. Wavelength calibration was performed using a set of HeNeArCd lamps observations obtained each afternoon and then adjusted using the location of 6 night sky lines in each science exposure (to account for flexure). We note that a stacked version of all exposures obtained on October 24 was first presented in \citetalias{Grondin2024} to confirm that Alessi12-PCE is a WD+MS binary. This spectrum showed many of the same features observed in the Keck and Gemini spectra described above, albeit at lower signal-to-noise. In this manuscript, we use the individual Kast exposures to measure RVs for Alessi12-PCE (see Section~\ref{sec:orbits}).

\subsection{Broadband Photometry}

In addition to spectroscopy, we produce an ultraviolet-optical-infrared spectral energy distribution for the Alessi12-PCE system from a range of archival and new broadband photometry.

\subsubsection{Archival Pan-STARRS1 and 2MASS Photometry}\label{sec:broadband}
We first query both the Panoramic Survey Telescope and Rapid Response System \citep[Pan-STARRS1;][]{2012ApJ...750...99T, 2016arXiv161205560C} and the Two Micron All Sky Survey \citep[2MASS;][]{2006AJ....131.1163S} at the position of the Alessi12-PCE system. From Pan-STARRS1 we retrieve photometry in the $g$, $r$, $i$, $z$, and $y$ bands (spanning effective wavelengths of $\sim4500-9500$~\AA). Specifically, we take the ``MeanPSF'' magnitudes for each filter. Magnitudes range from $\sim$18.8 AB mag ($g$-band) to $\sim$17.4 AB mag ($y$-band), with a typical standard deviation between measurements in a given filter of $\lesssim$0.035 mag. Between 10 and 14 observations were averaged per filter.  

From 2MASS, we retrieved $J$, $H$, and $K_s$ band photometry (spanning effective wavelengths of $\sim$1.2-2.2 $\mu$m). Magnitudes range from 16.2 Vega mag ($J$-band) to 15.2 Vega mag ($K_s$-band). We note that there are no other optical sources within 3$\arcsec$ of Alessi12-PCE in the Pan-STARRS1 catalog, so blending of multiple sources in the 2MASS catalog is not expected to be an issue. The Pan-STARRS1 and 2MASS photometry is shown in the main panel of Figure~\ref{fig:observedWDspectrum}. These reveal that the red component of the Alessi12-PCE spectrum peaks in flux between $\sim$8500-9500\AA.

\subsubsection{Swift/UVOT Photometry}

In addition to the archival photometry, we obtained UV observations of the Alessi12-PCE system with the Ultraviolet Optical Telescope (UVOT) onboard the \emph{Swift} satellite \citep{Gehrels2004, Roming2005}. Images were obtained in all six filters (UVW2, UVM2, UVW1, U, B, V; spanning effective wavelengths of $\sim$1900-5500 \AA). We retrieved the level 2 \emph{Swift-}UVOT sky images from the UK data archive\footnote{\url{https://www.swift.ac.uk/archive/browsedata.php?oid=00018982001&source=obs&reproc=1}}. We then calculated photometry for the Alessi12-PCE system using the standard \texttt{uvotsource} routine distributed with HEASARC. We adopted a source aperture with a radius of 5$\arcsec$ centered on the coordinates of Alessi12-PCE and a circular background region with a radius of 30$\arcsec$ (centered at $\alpha$ = 20:46:15.92 and $\delta$ = +24:38:31.96). We calculated photometry for the source in both individual exposures for each filter, as well as for a stacked image, produced with the HEASARC task \texttt{uvotimsum}. The source is detected at greater than 3$\sigma$ in the stacked images for all filters except for the V-band.

The resulting photometry is presented (on the AB magnitude scale) in Table~\ref{tab:uvot-obs}. In addition, we plot the UV, U, and B-band results from the stacked images in the main panel of Figure~\ref{fig:observedWDspectrum} (we do not plot the V-band result as it is only an upper limit and overlaps in wavelength with Pan-STARRS). This photometry highlights that the flux of the blue component in the Alessi12-PCE system continues to rise in the UV and therefore must peak at a wavelength shorter than $\sim$2000 \AA.\footnote{We note that the photometry plotted in Figure~\ref{fig:observedWDspectrum} has not been corrected for interstellar extinction.}

\subsection{Zwicky Transient Facility Light Curve} \label{sec:lightcurvesalessi12}

As in \citetalias{Grondin2024}, we retrieve the light curve of Alessi12-PCE from the Zwicky Transient Facility \citep[ZTF;][]{Masci2019}. We utilized the NASA/IPAC Infrared Science Archive\footnote{\url{https://irsa.ipac.caltech.edu/docs/program_interface/ztf_lightcurve_api.html}} to query all $g$- and $r$-band light curves located within a 5$\arcsec$ radius of the position of Alessi12-PCE ($\alpha=311.604^{\circ}, \delta= 24.638^{\circ}$). Next, we remove any duplicate observations or epochs that were flagged as poor observations (\texttt{catflags > 32768}). This filtering leaves 919 high-quality r-band observations over a $\sim 4.7$ year time period in the ZTF catalog. 

\section{Spectral Fitting} \label{sec:analysis}
We jointly fit a combined WD+M dwarf model to determine the individual parameters for both the WD and M dwarf in Alessi12-PCE. For this analysis, we focus on fitting our Keck/LRIS spectrum (Figure \ref{fig:observedWDspectrum}) as it is our highest SNR measurement. Below, we outline the WD and M dwarf models used in the fitting (Section \ref{sec:wdmodels}), along with a routine that first determines the spectral type of the M dwarf (Section \ref{sec:fluxfits}) and then compute the atmospheric parameters ($T_{\text{eff}}$ and $\log{g}$) of the WD (Section \ref{sec:wdmodelling}).

\subsection{WD and M Dwarf Models}\label{sec:wdmodels}
Our Keck spectrum reveals clear, broad higher-order Balmer absorption lines (H$\beta$, H$\gamma$, H$\delta$, and H$\epsilon$), indicating that the atmosphere of the WD is hydrogen-dominated (DA). As such, we use the non-local thermodynamic equilibrium (NLTE)  DA WD atmospheric models developed by \cite{Tremblay2011} to model the WD. By coarsely plotting a few example WD models against our observed spectrum, we define a rough parameter space for our WD models. Our final grid contains 212 WD models spanning $30{,}000 \leq T_{\mathrm{eff}}\,(\mathrm{K}) \leq 40{,}000$ and $8.0 \leq \log g\,(\mathrm{cm\,s^{-2}}) \leq 9.5$. 

To model the M dwarf, we use a set of \cite{Pickles1998} stellar models, which include M3V, M4V, and M5V spectral types. This yields 636 distinct WD+M dwarf model pairs that are fit to the observed spectrum. It is worth noting that there are some spectroscopic features observed that are not present in the models. In particular, as seen in Figure \ref{fig:observedWDspectrum}, Alessi12-PCE shows narrow Balmer emission lines inside broader absorption features. Such features can be formed in the chromosphere of both isolated M dwarfs \citep[e.g.,][]{2017ApJ...834...85N} and those in tight binaries \citep{Silvestri2006}.

\subsection{Relative Flux Scalings and M Dwarf Spectral Type}\label{sec:fluxfits}

\begin{figure*}
    \centering
    {\includegraphics[height=8.5cm]{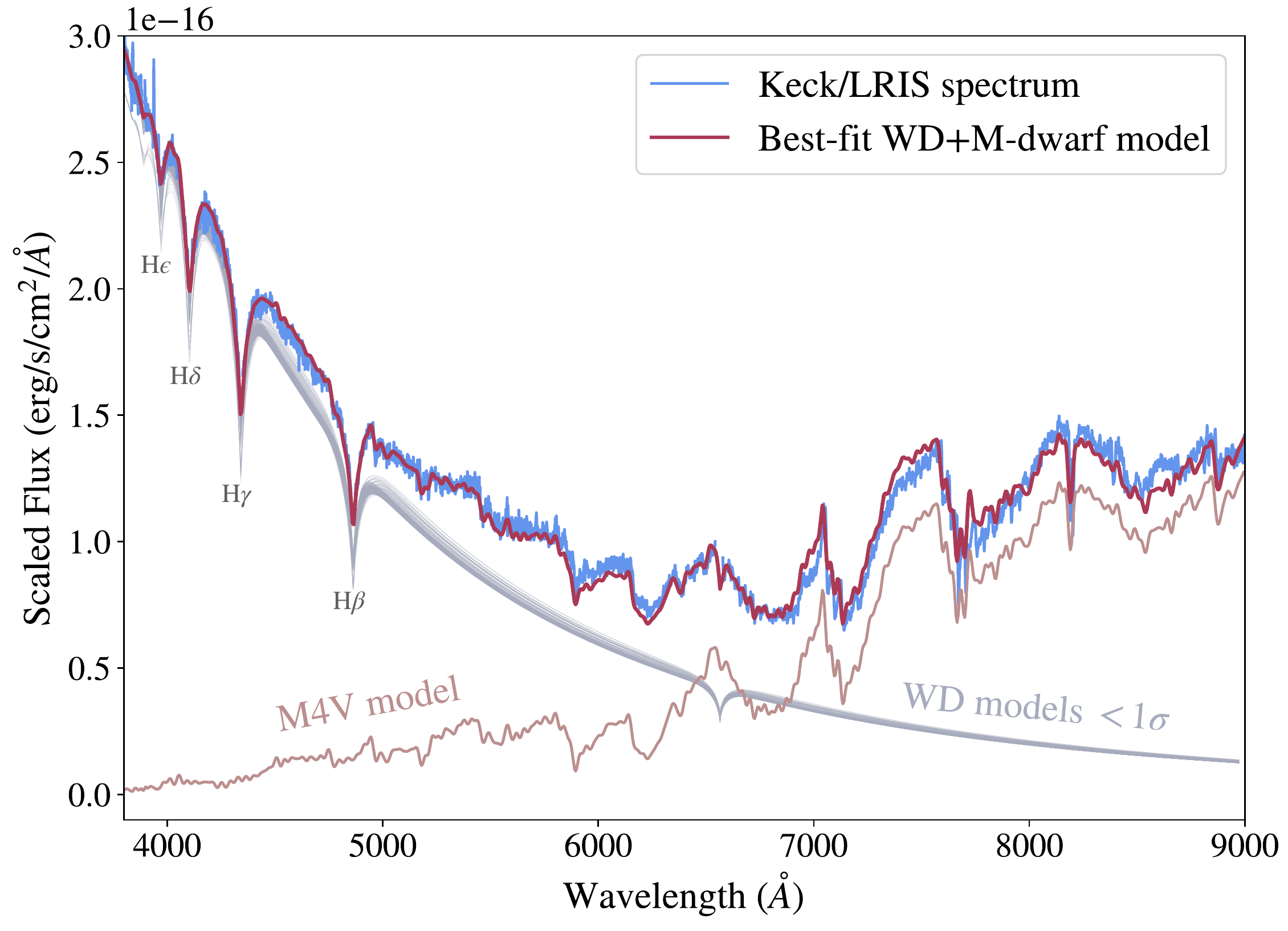}}
    \raisebox{-0.135cm}{
    \includegraphics[height=8.48cm]{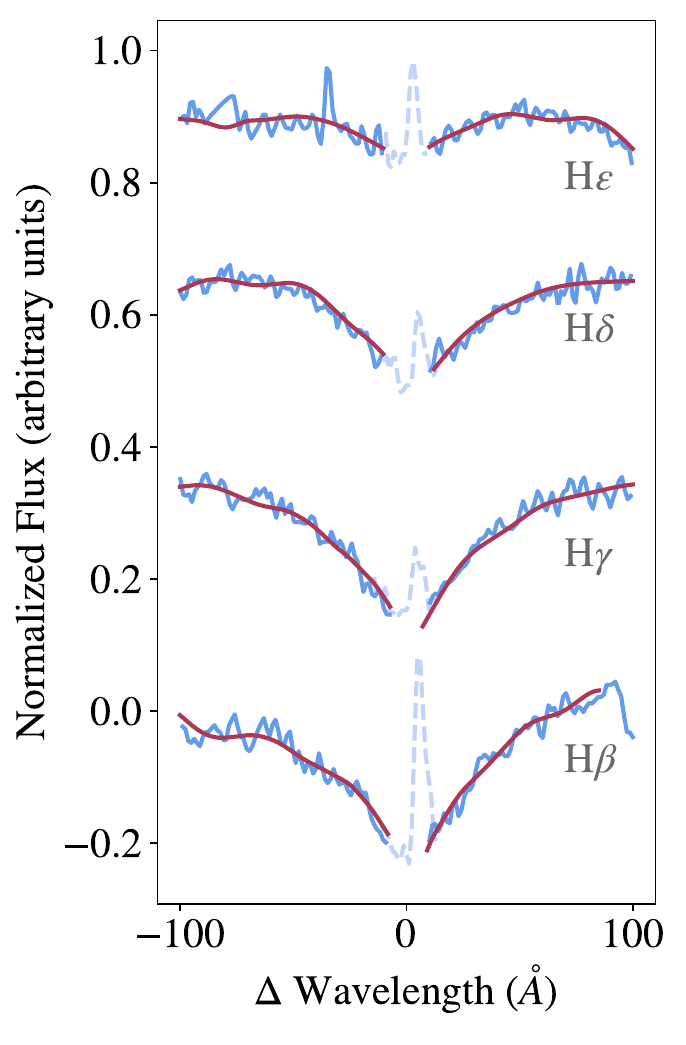}}
    \caption{Summary of fitting properties to determine the M dwarf spectral type and WD atmospheric parameters ($T_{\text{eff}}$ and $\log{g}$). \textit{Left:} Flux-space spectral fitting of Alessi12-PCE. Although a broad range of WD atmospheric models provide acceptable fits to the observed continuum (gray), only models containing an M4V companion are consistent within $1\sigma$, demonstrating that the companion's spectral type is robustly constrained.
    The best-fitting combined WD+M4V model is shown in pink and the observed Keck/LRIS spectrum in blue. \textit{Right:} Normalized flux spectral fitting of Alessi12-PCE. Specifically, we fit the higher-order Balmer absorption lines of H$\beta$, H$\gamma$, H$\delta$, and H$\epsilon$. The best-fit model (pink) is a WD with $T_{\text{eff}}=39,250$K and $\log{g}=8.70$. Note that we masked the emission lines present in the cores of each Balmer absorption feature, as these lines are not included in the spectral models. We show them here (blue dashed lines) for reference.}
    \label{fig:fitting}
\end{figure*}

We first fit the Keck spectrum in flux space, using only data $3800 \leq \lambda\,(\mathrm{\AA}) \leq 9000$ as the spectrum is dominated by noise outside this range.
We first scale the flux of the spectrum to match the photometry in the Pan-STARRS1 $g$-band 
and then correct the spectrum for reddening, adopting an extinction value of $A_{V} = 0.10$ measured for the Alessi 12 star cluster \citep{CG2020} and a \cite{Cardelli1989}
reddening law with a standard $R_{V} = 3.1$ for the Milky Way. 
We mask out a $\pm 10-15 $ \AA~region around the central wavelengths of the H$\beta$, H$\gamma$, H$\delta$, and H$\epsilon$ Balmer absorption lines, to exclude the emission features present in the spectrum, since these features are not included in our WD or M dwarf model spectra. We note, however, that the presence of these features does not impact our model fits.

For each WD and M dwarf model combination, we first determine the relative flux scaling between the two stars that minimizes $\chi^{2}$. To achieve this, we calculate the normalization required for the WD model to match the observed flux from $\lambda = 3800-3850$  \AA~and for the M dwarf model to match the observed flux from $\lambda = 8800-8850$ \AA. These wavelength ranges were selected because they are free of strong spectral features and lie sufficiently far in the blue and red that the WD and M dwarf are respectively expected to dominate the observed flux. We create a grid of combined WD+M dwarf models where we (independently) vary the flux normalization of both components between 0.5 and 1.5 times the `base' flux normalization described above. We then determine which combination of flux scalings yield the best-fit to the observed data for each particular WD+M dwarf combination, for all models in our grid.

Our primary goal when performing this flux-space fit is to determine the M dwarf spectral type, while the WD properties ($T_{\text{eff}}$ and $\log{g}$) will be further examined in Section \ref{sec:wdmodelling} (mainly due to the potential impact of flux calibration uncertainties). Overall, we find that every WD+MS pair that was allowed within a $1\sigma$ confidence interval used an M4V MS model. We therefore conclude that the companion in Alessi12-PCE is most consistent with an M4V star\footnote{We note that this M dwarf spectral type (M4V) differs from the spectral type (M3V) plotted in the original spectrum of Alessi12-PCE in Figure 10 of \citetalias{Grondin2024}. The fit presented in this work is of much higher quality, due to the higher SNR spectrum of this source. Properties of the M dwarf will be further discussed in Section \ref{sec:mdwarfproperties}.}. The results from this flux-space spectral fitting are shown in the left panel of Figure \ref{fig:fitting}.
\subsection{WD Modeling}\label{sec:wdmodelling}

We determine the best-fitting WD parameters of Alessi12-PCE by fitting the higher-order Balmer absorption lines (H$\beta$, H$\gamma$, H$\delta$, and H$\epsilon$) using flux-normalized spectra for 192 WD+M4V models with $31,000 \leq T_{\mathrm{eff}}(\mathrm{K}) \leq 40,000$ and surface gravities of $8.5 \leq \log g (\mathrm{cm,s^{-2}}) \leq 9.5$ that were allowed within $1\sigma$ in the preliminary fits described in Section \ref{sec:fluxfits}\footnote{The WD models allowed within this range span effective temperatures of $31,000 \leq T_{\mathrm{eff}}(\mathrm{K}) \leq 40,000$ and surface gravities of $8.5 \leq \log g (\mathrm{cm,s^{-2}}) \leq 9.5$.}. 
The width and depth of the lines (i) contain information on the effective temperature and surface gravity of the WD and (ii) are less sensitive to continuum shape uncertainties due to flux calibration/reddening when determining WD temperature. Using the relative flux scalings between the WD and M4V stars from Section \ref{sec:fluxfits}, we first trim each combined model spectrum to $3800–5200$ \AA~and then normalize both the observed spectrum and model grid in \texttt{pyraf} with the task \texttt{continuum}, employing a cubic spline of order 7. To determine the best fit $T_{\text{eff}}$ and $\log{g}$, we again compute a $\chi^{2}$ statistic according to the same method as our flux-space analysis in Section \ref{sec:fluxfits}. The right panel of Figure \ref{fig:fitting} highlights the results of this spectral fitting in normalized space, where the best-fit WD model is a system with $T_{\text{eff}}=39,250$K and $\log{g}=8.70$. We also show the emission features that were masked in the spectral fitting routine.

To estimate uncertainties, we use Monte Carlo resampling. Specifically, we generate 1,000 realizations of our observed spectrum by sampling each pixel from a normal distribution described by its value and measurement error. We then fit our full grid of WD+M4V models to each realization. From the distribution of best-fit parameters obtained from our 1,000 spectral realizations, we adopt the mean and standard deviation as our final WD parameter estimates. We find that the WD in Alessi12-PCE has an effective temperature of $T_{\mathrm{eff}} = 39{,}100 \pm 300~\mathrm{K}$ and a surface gravity of $\log g = 8.69 \pm 0.04$ dex. 

\section{Orbital Parameters} \label{sec:orbits}

\citetalias{Grondin2024} found that Alessi12-PCE exhibits clear and regular variability in a ZTF light curve (described in Section \ref{sec:lightcurvesalessi12}).
Here, we analyze the light curve to determine a periodicity and present RV measurements of both H$\alpha$ emission features and short-wavelength Balmer absorption lines from our spectroscopy described in Section \ref{sec:obs}. The main outcome from this section that will be used in our subsequent analysis is a measurement of the current orbital period of the binary system.

\subsection{Light Curve Periodicity}\label{sec:lightcurve}

In post-CE WD+MS systems, variability is often observed due to (i) eclipses,  (ii) ellipsoidal modulations, (iii) reflection effects, or (iv) M dwarf rotation/starspots \citep[][]{NebotGomezMoran2011}. Binary eclipses can place strong constraints on the orbital period and inclination, but are rare.  Ellipsoidal modulations occur when one (or both) stars become slightly elongated (``ellipsoidal" in shape rather than spherical) due to tidal distortions from the companion star. Reflection effects occur when light from the primary star (in this case, the WD) irradiates the surface of its companion (the M dwarf) causing changes in brightness throughout its orbital period. Without RV monitoring, it is difficult to determine the mechanism driving variability in the system. However, light curves can help constrain the period of a binary, making them useful tools for characterizing post-CE binaries.

\begin{figure*}
    \centering
    \includegraphics[width=\textwidth]{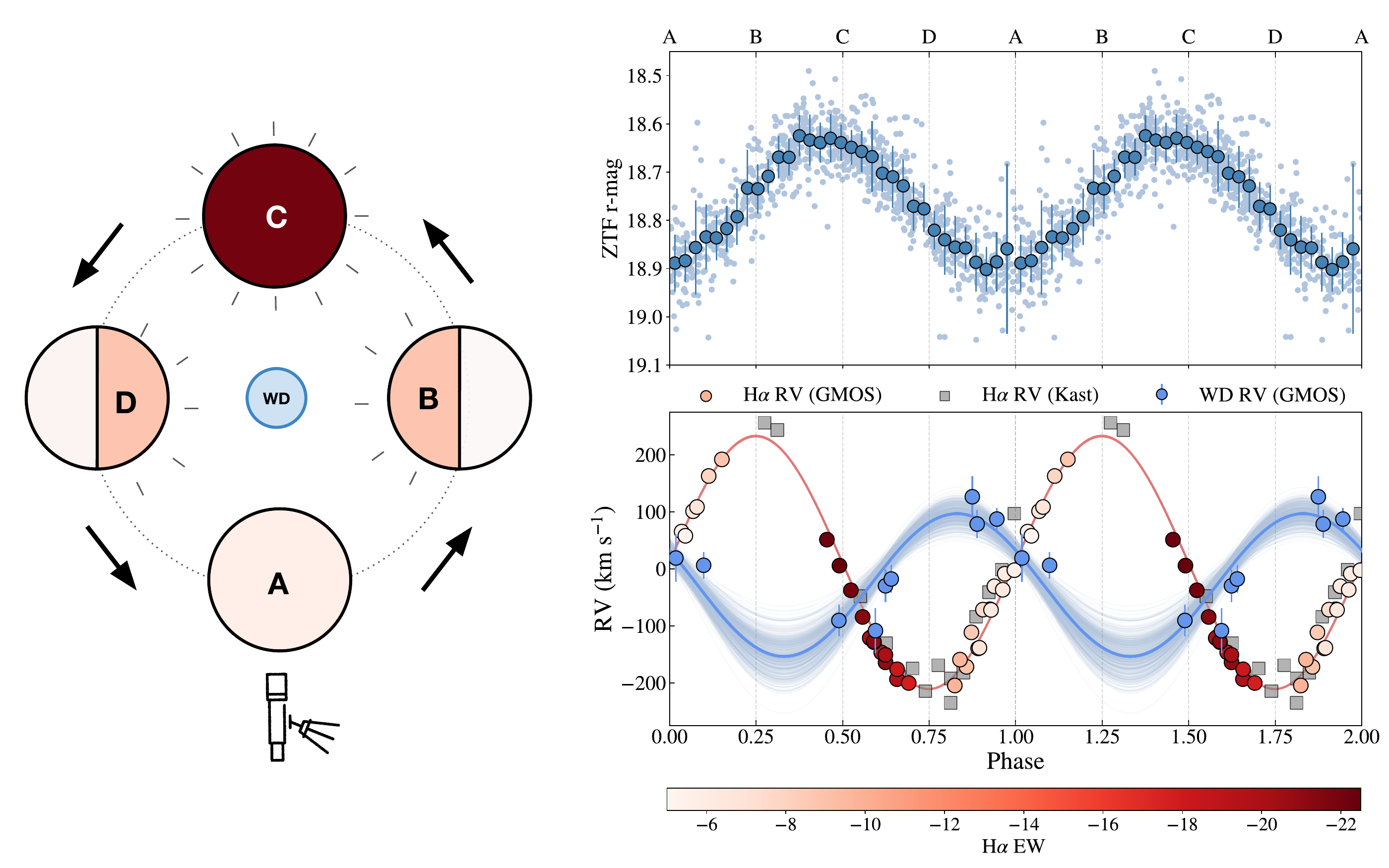}
    \caption{The orbital variability of Alessi12-PCE. \textit{Left panel:} A schematic illustrating variability due to a reflection/irradiation effect. Here, the WD irradiates one hemisphere of the tidally locked M dwarf, so the same face always points towards the WD throughout the binary's orbit. An orbital phase of 0 (`A') corresponds to the time when the irradiated face is hidden from the observer (minimum observable irradiation), whereas an orbital phase of 0.5 (`C') is when the M dwarf's irradiated face is completely in view (maximum observable irradiation). Orbital phases of 0.25 (`B') and 0.75 (`D') are where the irradiated regions are partially visible. \textit{Top right panel:} 
   A phase-folded r-band ZTF light curve of Alessi12-PCE reveals a periodicity of $P=6.99$ hours. We bin the magnitudes over every 0.03 in phase to highlight the system's general behavior. The light curve is faintest at a phase of 0 and brightest at a phase of 0.5, which aligns with the reflection effect scenario in (1). \textit{Bottom right panel:}  The phase-folded RVs measured from the H$\alpha$ Balmer emission line are shown for 31 epochs of Gemini/GMOS spectroscopy (red circles, color-coded by H$\alpha$ emission width and 16 epochs of Shane/Kast spectroscopy (gray squares). The RVs measured from the H$\beta$ Balmer absorption line from 9 epochs of Gemini/GMOS spectroscopy are indicated as blue circles, where different fits from a Monte Carlo analysis of 500 iterations are shown. All data are folded on the light curve period from above. The H$\alpha$ emission line velocities trace the motion of the M dwarf companion, while the H$\beta$ absorption line velocities trace the motion of the WD. These measurements confirm that this period corresponds to the orbital period of Alessi12-PCE. The RVs reach the systematic velocity of the system at orbital phases of 0 and 0.5, again well-matching the reflection effect picture. The H$\alpha$ RVs are colored by their emission line width strengths, which are strongest at an orbital phase of 0 and weakest at an orbital phase of 0.5, indicating that the H$\alpha$ emission lines are likely coming from the irradiated M dwarf in the system.}
    \label{fig:alessi12variability}
\end{figure*}

To determine a periodicity from the light curve, we compute a Lomb-Scargle periodogram \citep{Lomb1976, Scargle1982} using the \texttt{LombScargle} function from \texttt{astropy.stats} \citep{2013A&A...558A..33A}. We test a period range of 0.25-24 days and find a maximum power at $P=0.2916 \ \text{days}=6.9984 \ \text{hours}$ in the periodogram. The periodogram and ZTF phase folded light curve are plotted in Figure \ref{fig:alessi12variability}. 
To estimate the uncertainty on this peak period, we again perform a Monte Carlo analysis. Each realization is analyzed with a Lomb-Scargle periodogram, where the distribution of resulting periods provides an estimate of the period uncertainty. From this analysis, we find that the uncertainty on the measured variability period due to photometric errors is very small: $P=0.29163810 \pm 0.00000033$ days. It is worth noting that this uncertainty only includes noise about the peak identified period, where aliasing/systematic uncertainties from ZTF may also be present. The other frequencies in the periodogram are harmonics of this period and we do not see any evidence for eclipses in the ZTF light curve. 

\subsection{Radial Velocities}

\subsubsection{H$\alpha$ Emission Line RVs}

As illustrated in Figures \ref{fig:observedWDspectrum} and \ref{fig:gemini-allspec}, the H$\alpha$ Balmer line exhibits a prominent emission feature. In all individual spectroscopic exposures (Keck, Gemini, and Shane) we detect the H$\alpha$ line in emission. Although the strength of the emission varies between observations, this feature remains consistently visible in each spectrum. Many post-CE WD+M dwarf binaries exhibit H$\alpha$ emission features \citep{RM2010, 2013MNRAS.429.3570R}, which can, in principle, be due to either the WD or M dwarf. For instance, isolated DA(e) WDs have been observed to exhibit H$\alpha$ in emission due to magnetic mechanisms \citep{2023MNRAS.524.4996E}. On the other hand, low-mass M dwarfs are highly (or
even fully) convective, which can generate large magnetic fields in stellar chromospheres and cause
stellar activity in the form of coronal mass ejections, eruptions or flares \citep[e.g.,][]{2024ARA&A..62..593H}. For M dwarfs with close companions, tidal locking synchronizes their rotation period with the orbital period of the binary. This rapid rotation can further amplify magnetic fields, increasing the chromospheric activity in M dwarfs even more in these close binary systems \citep{2012AJ....144...93M}.

Regardless of its physical origin, RVs of the H$\alpha$ features can be used to measure the orbital period of the system, and test its alignment with the variability timescale measured from the ZTF light curve. To derive the RVs based on the H$\alpha$ emission feature, we first fit a pseudo-continuum to the 6250--6850 \AA\ region of each of the 33 Gemini and 14 Shane spectra\footnote{While there are 18 individual Shane/Kast exposures with the red detector (which cover the H$_\alpha$ feature) in Table~\ref{tab:obssummary}, we found that four exposures lacked sufficient signal-to-noise for analysis (two on October 24, 2023 and two on December 5, 2023).}  (i.e., each individual exposure) with a low-order polynomial. We then fit a Gaussian to the H$\alpha$ emission line in each spectrum using the \texttt{scipy curve\_fit} package. From this, we measure both the central wavelength as well as the equivalent width of the emission line in each spectrum. Assuming a rest frame H$\alpha$ central wavelength of $\lambda_{\text{rest}} = 6562.8$ \AA~and using the equation for relativistic Doppler shift, we convert each $\lambda_{\rm obs}$ into an RV. To estimate uncertainties, we again run a Monte Carlo simulation (as in Sections \ref{sec:wdmodelling} and \ref{sec:lightcurve}), where we sample each data point from a normal error distribution 100 times and repeat the analysis above. We take the means and standard deviations of these iterations as our final H$\alpha$ RVs and associated errors.

Figure \ref{fig:alessi12variability} highlights the combined H$\alpha$ RVs from Gemini/GMOS and Shane/Kast phase-folded by the previously determined ZTF period of $P=0.2916$ days. From this, we see that the variability timescale observed in the ZTF light curve corresponds to the orbital period of Alessi12-PCE, but offset by a quarter phase (i.e., the peaks/troughs of the RV curve correspond to the mid-points of the brightness distribution in the light curve). Fitting a sine function to the phase-folded H$\alpha$ RV data yields an observed semi-amplitude $K_{\rm{H}\alpha}=221.86 \pm 0.34$ km~$\rm s^{-1}$ and a systemic velocity of $v_{\rm{sys}}= 10.97\pm 0.29$ km $\rm s^{-1}$. This fit is shown in Figure~\ref{fig:alessi12variability}, where the dark line is the best-fit and lighter pink lines correspond to results from a Monte Carlo analysis where the RVs are varied based on their measured errors and the fit is repeated.

\subsubsection{WD RVs}

As illustrated in Figure \ref{fig:fitting}, the H$\beta$ and H$\delta$ Balmer absorption lines are dominated by the WD. We can therefore measure RVs from the Balmer absorption features to determine whether we observe anti-correlated motion relative to the M dwarf H$\alpha$ RVs, described above.

Measuring precise RVs from the Balmer absorption features is more challenging than the M dwarf H$\alpha$ emission line feature since they are comparatively weaker, and broader. As such, we utilize the 9 higher SNR stacks of our Gemini data described in Section~\ref{sec:geminispec} and shown in Figure~\ref{fig:gemini-allspec} to measure the WD RVs\footnote{We note that the individual Shane/Kast data are not sufficiently high SNR to measure WD RVs, even if stacking all observations obtained on a given night.}. Each of these consist of 2-4 individual exposures taken sequentially. 

For each of these 9 spectra, we first fit a pseudo-continuum to the 3750-5200 \AA~region to normalize. We then undertake a two step process to determine the WD RVs. First, we take the best-fit WD template from Section~\ref{sec:wdmodelling} and determine the RV shift that minimizes $\chi^2$ with respect to the observed H$\beta$ and H$\delta$ (with fit windows of 70 and 120 \AA, respectively). When performing this fit we mask both the centers of the lines (which are contaminated by emission) and the red wing of the H$\beta$ line (which contains a M dwarf feature). After performing this preliminary analysis, we find that the WD RVs are anti-correlated with the H$\alpha$ RVs measured in the previous section, indicating that this is a double-lined spectroscopic binary (SB2). This anti-correlated motion indicates that the H$\alpha$ emission feature originates predominately from the M dwarf, and we use this to refine our WD RV analysis by taking into account contributions from the M dwarf in the regions of the spectra overlapping H$\beta$ and H$\gamma$.

Since the binary is an SB2, we measure the RVs of the WD from H$\beta$ and H$\gamma$ while accounting for the RV motion of the M dwarf. First, we determine the relative flux contribution of both the M dwarf and the WD from the best-fit models, and dilute both templates accordingly (see Section \ref{sec:analysis}). Second, we shift the M dwarf template by the RV inferred from the best-fit H$\alpha$ orbit (see above). Third, we fit the WD RVs for each stacked spectrum by minimizing $\chi^2$ with respect to the RV shift of the best-fit WD template relative to the observed H$\beta$ and H$\delta$ (with fit windows of 70 and 120 \AA, respectively). We again estimate uncertainties through a Monte Carlo simulation, sampling 1,000 times and repeating the analysis above. We take the final joint H$\beta$ and H$\delta$ RVs to be the means and standard deviations of these iterations. 

Figure \ref{fig:alessi12variability} highlights the phase-folded H$\beta$ and H$\delta$ RVs (blue), relative to the H$\alpha$ RVs (red). Interestingly, when fitting a sine function to the (small number of) WD RVs, we find a small phase offset of $\lesssim$0.1 between the M dwarf and WD RV curves (as above, the dark blue line in Figure~\ref{fig:alessi12variability} shows the best-fit curve, while lighter blue lines show results when varying each RV point based on its error). The origin of this phase offset is unclear, but one potential explanation is that the H$\alpha$ emission line is not tracing the M dwarf center of mass. This effect is established in dwarf novae: the emission line RV curve in a cataclysmic variable can have a larger amplitude and phase offset relative to the true orbital motion of the WD due to the presence of a disk \citep{1988ApJ...324..411W}. In our case, the H$\alpha$ emission likely originates on the irradiated surface of the M dwarf (see below), which is expected to trace the center of light, rather than the center of mass, of the M dwarf \citep{Parsons2010}. This could potentially induce a phase offset in the RVs that is either physical (e.g., if there are spots on the irradiated surface that impact the center of light) or the result of systematic errors\footnote{In particular, we highlight that our final RV measurements were calculated after shifting the best-fit M dwarf template based on the H$\alpha$ RVs. If the true semi-amplitude of the M dwarf center-of-mass is larger than that measured by H$\alpha$ this could induce a systematic error in some of our WD RV measurements. Although we note that RVs measured this way were very similar to those that did not account for the M dwarf in this way}. 

Regardless of its cause, we emphasize that the presence of this apparent phase offset does not impact any of the further analysis in this manuscript. The purpose of this section is to provide further insight into the origin of the H${\alpha}$ emission and this offset does not affect our conclusion that the Balmer absorption and H$\alpha$ emission features exhibit anti-correlated motion. In particular, the WD RVs vary by several hundred km s$^{-1}$, leaving little ambiguity in the detection of the companion’s anti-correlated orbital motion.

\subsection{The Orbital Configuration of Alessi12-PCE}\label{sec:reflection}

Results from the previous subsections demonstrate that Alessi12-PCE is a WD+M dwarf binary in a $P_\text{orb} \approx 7$ hour orbit. Given the alignment between the photometric variability and H$\alpha$ RVs, for our analysis below we take the orbital period to be that measured from the ZTF light curve. This short period is consistent with the population of post-CE binaries studied to date \citep[e.g.,][]{NebotGomezMoran2011} and we therefore proceed in the following sections to constrain both the pre- and post-CE properties of the binary system and use them to estimate $\alpha_{\text{CE}}$.    

Given that the orbital period is equal to the variation timescale observed in the ZTF light curve, we find that the origin of the variability cannot be ellipsoidal modulations (which have timescales corresponding to half the period), but more likely corresponds to physical processes such as a reflection effect. In particular, we also find that the strength of the H$\alpha$ emission line is also correlated with the orbital period. This finding is demonstrated in Figure~\ref{fig:alessi12variability} where we color code the H$\alpha$ measurements from the Gemini spectra by the equivalent width of the emission line (we do not color code the Shane measurements which are lower SNR).

We see a clear trend with orbital phase where the strongest (weakest) emission lines are observed in conjunction with the brightest (dimmest) portions of the light curve (at phases of 0.5 and 0.0, respectively). This matches expectations for systems where both the variability and emission lines are caused by an irradiation/reflection effect on the surface of the M dwarf star. In this case, the Balmer emission lines would originate from the M dwarf and phase = 0.5 would correspond to when the irradiated face of the M dwarf is maximally visible (see the schematic in Figure \ref{fig:alessi12variability}). We note that in this case, the H$\alpha$ emission feature originates from the irradiated face of the M dwarf rather than its center of mass. As a result, the semi-amplitude inferred from the H$\alpha$ RVs is expected to underestimate the true value \citep{Parsons2010}. 

\section{Post-CE Properties}\label{sec:masses}
In order to link the pre- and post-CE conditions of Alessi12-PCE, we must first determine (i) the current masses of both the M dwarf and WD in the system and (ii) the current separation of the binary.

\begin{figure*}[!ht]
    \centering
    \includegraphics[width=1\linewidth]{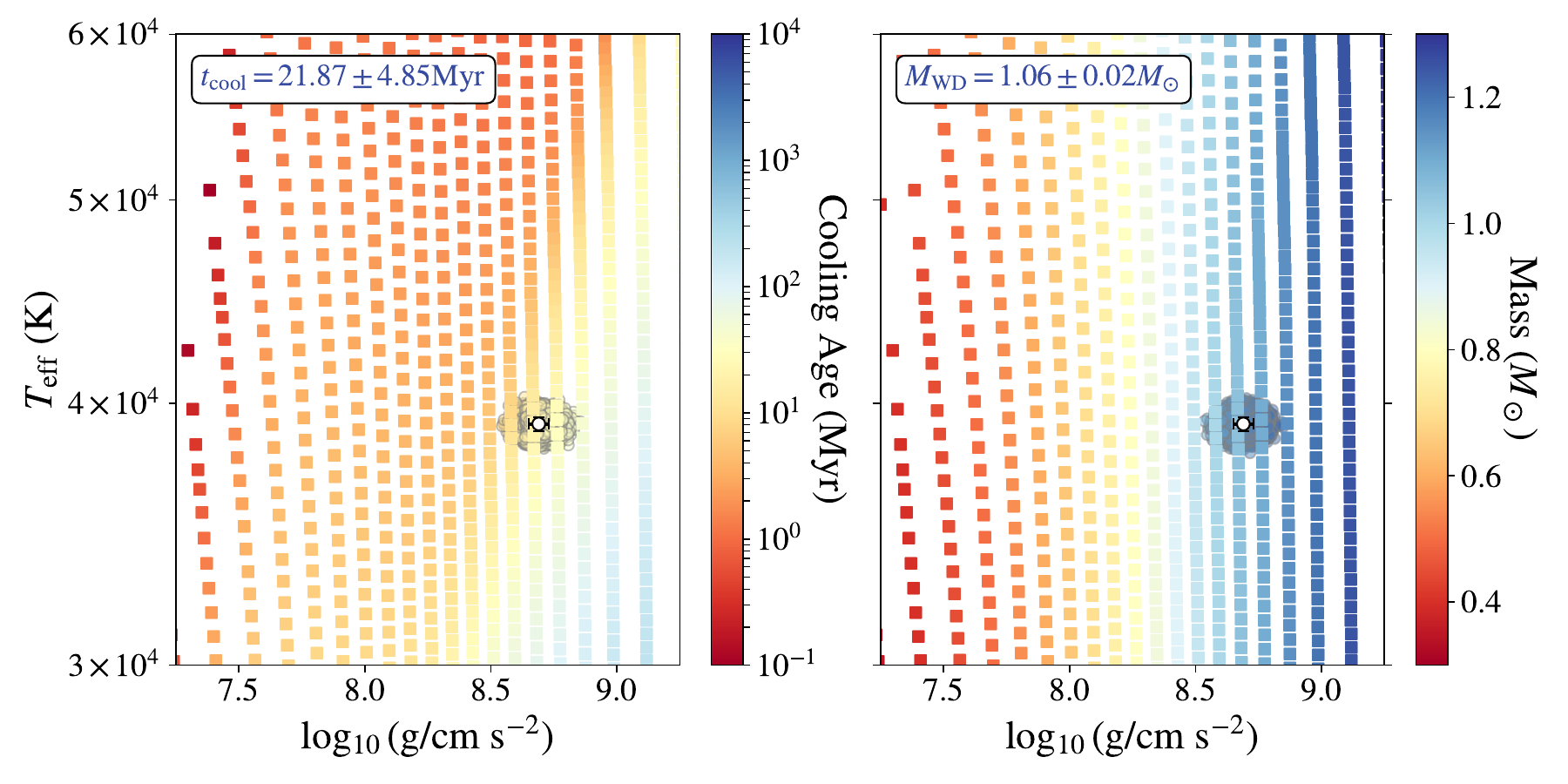}
    \caption{A Monte Carlo analysis to determine the cooling age ($t_{\rm{cool}}$; left plot) and mass ($M_{\rm{WD}}$; right plot) of the WD in Alessi12-PCE, using the \cite{2020ApJ...901...93B} ``thick" DA WD atmosphere models (colored grid). All Monte Carlo results are plotted as small circles and our previously derived WD $T_{\text{eff}}$ and $\log{g}$ are indicated as white circles, indicating the derived WD $t_{\rm{cool}}$ and $M_{\rm{WD}}$.}
    \label{fig:wdcooling}
\end{figure*}

\subsection{M Dwarf Mass}\label{sec:mdwarfproperties}

To determine the current mass of the M dwarf, we use the mass-magnitude ($M_{*}-M_{K_{S}}$) relation presented in \cite{Mann2019}. We determine the M dwarf's absolute $K_{s}$-band magnitude using its apparent 2MASS $K_{s}$ magnitude \citep[$m_{K_{s}}=15.23 \pm 0.16$ mag;][]{2006AJ....131.1163S}, the mean distance to the star \citep[$d=566^{+50}_{-44}$ pc;][]{2021AJ....161..147B}, and the $K_{s}$-band extinction. The extinction is computed as $A_{K_{s}}= \frac{A_{V}}{R_{V}}\cdot k(K_{s})$, where we adopt the V-band extinction of the Alessi 12 star cluster \citep[$A_{V}=0.10$;][]{CG2020}, the standard extinction ratio for the Milky Way \citep[$R_{v}=3.1$;][]{Cardelli1989}, and the 2MASS  $K_{s}$-band extinction coefficient \citep[$k(K_{s}) = 0.306$;][]{Yuan2013}. By propagating errors from both the distance and magnitude we find an absolute magnitude of $M_{K_{s}} = 6.46^{+0.25}_{-0.23}$ mag. Adopting the preferred $n=5$ fit and the corresponding coefficients in Table 6 of \cite{Mann2019}, we determine $M_{*}$ by randomly sampling 1000 $M_{K_{s}}$ values from a normal distribution with a mean of 6.46 mag and standard deviation of 0.24 mag. From this analysis, we measure a mass of $M_{*}=0.370 \pm 0.044\, M_{\odot}$ for the M4V companion in Alessi12-PCE. 

It is worth noting that there is known scatter in the mass-radius relationship of close M dwarf binaries (including those containing WDs), where radii can be inflated by up to 12\% \citep{2018MNRAS.481.1083P}. Such an effect may explain the slight mismatch between the mass we infer and those typically associated with M4V stars (our mass most closely matches those associated with M3V stars in Table 7 of \citealt{Mann2019}).
By comparing dynamical mass measurements to those inferred from the \cite{Mann2019} relation, \cite{2018MNRAS.481.1083P} showed that the $M_{}$–$M_{K_s}$ relation may systematically underpredict M dwarf masses in tight binaries by $\sim$5–10\%. To account for this, we include a 10\% systematic uncertainty in quadrature with our statistical error, resulting in a final M4V mass estimate of $M_{*} = 0.370 \pm 0.058 M_{\odot}$.

\subsection{White Dwarf Mass and Cooling Age}\label{sec:wdmasscooling}

We determine a current WD mass based on the $T_{\text{eff}}$ and $\log{g}$ measured in Section~\ref{sec:wdmodelling}. To do so, we use a model grid of WD atmospheres from \cite{2020ApJ...901...93B}\footnote{\href{https://www.astro.umontreal.ca/~bergeron/CoolingModels/}{https://www.astro.umontreal.ca/$\sim$bergeron/CoolingModels/}}. 
Different WD evolutionary pathways predict different hydrogen envelope compositions, where this suite of models provides two different hydrogen mass layers for the WD: a ``thick" \citep[$q_{H}=10^{-4}$;][]{1984ApJ...282..615I} and a ``thin" \citep[$q_{H}=10^{-10}$;][]{1987fbs..conf..319F} layer. Thin hydrogen atmospheres in WDs are often attributed to merger events that partially strip the outer layers of the star \citep[e.g.][]{2025NatAs...9.1347S}. Because the WD properties and age of the Alessi 12 star cluster rule out a merger scenario (see Section \ref{sec:mergerproduct}), we adopt a thick hydrogen atmosphere model in our analysis.

\begin{figure*}
    \centering
    \includegraphics[width=1.0\linewidth]{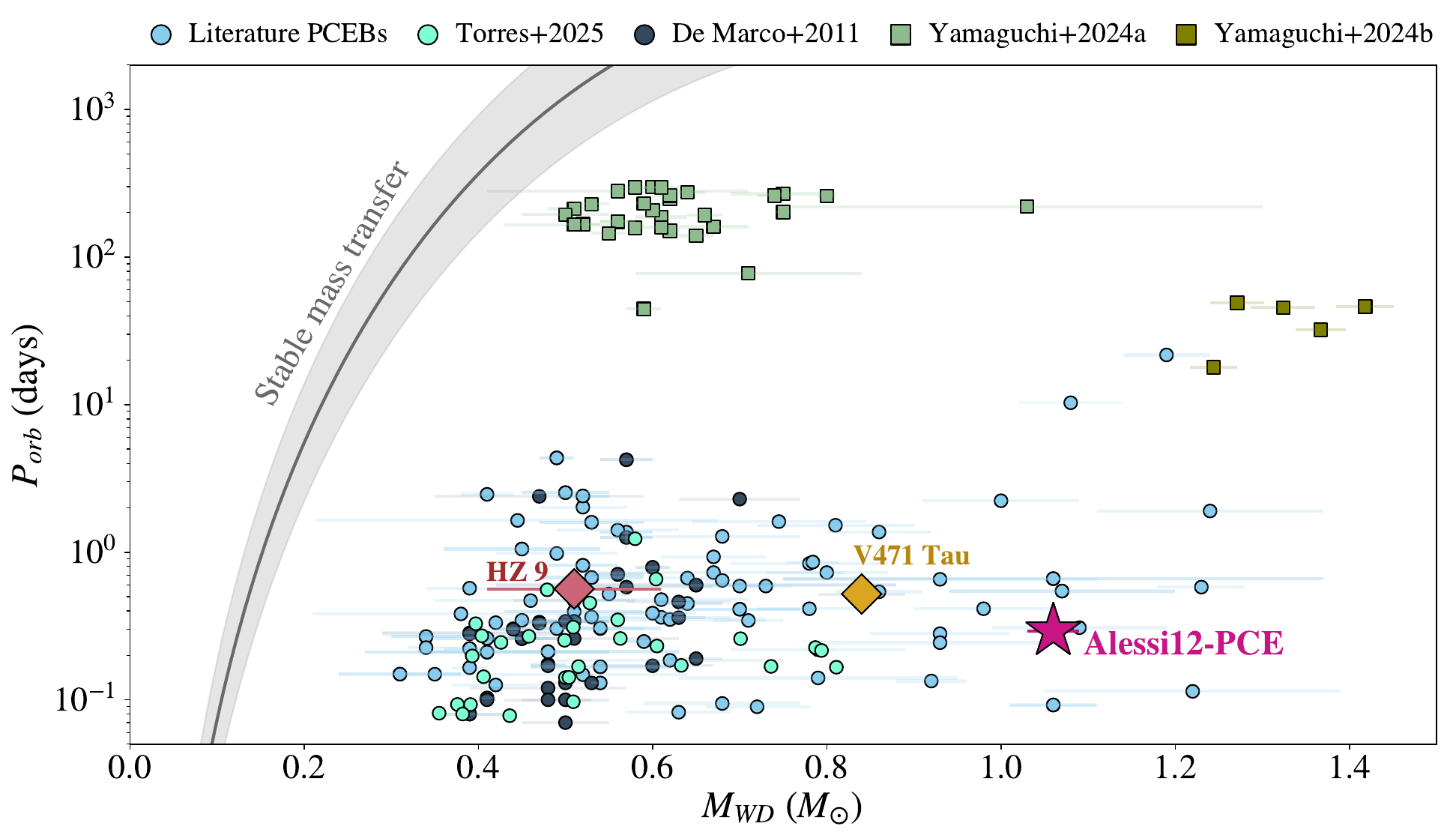}
    \caption{Location of Alessi12-PCE (pink star) in the $P_{\rm orb}$–$M_{\rm WD}$ plane,
compared with the currently known population of detached WD+MS post-CE binaries. With
$M_{\rm WD} = 1.06 \pm 0.02\,M_{\odot}$, Alessi12-PCE is the most massive such system
associated with an open star cluster. Comparison systems are grouped by the role they play.
\textit{Cluster systems (diamonds):} the only two other detached WD+MS post-CE binaries
known in open clusters, V471 Tau (gold) and HZ~9 (coral). \textit{Systems with reconstructed pre-CE parameters
(circles):} \cite{2025A&A...698A.173T} (cyan) and \cite{2011MNRAS.411.2277D} (dark blue),
which derive CE efficiencies ($\alpha_{\rm CE}$) for their post-CE samples in a manner
similar to this work. \textit{Long-period samples (squares):} \cite{2024PASP..136h4202Y}
(light green) and \cite{2024MNRAS.52711719Y} (olive). \textit{Broader field population
(cornflower blue):} additional systems from \cite{Zorotovic2010}, \cite{NebotGomezMoran2011},
\cite{2021MNRAS.501.1677H}, \cite{2022MNRAS.517.2867H}, \cite{2022MNRAS.512.1843H}, and
\cite{2026AJ....171..159M}. The dashed gray line marks the predicted boundary for stable
mass transfer \citep{1995MNRAS.273..731R}. This figure is adapted from
\citet{2024PASP..136h4202Y}.}
    \label{fig:summarysystems}
\end{figure*}

Using the thick set of WD atmospheric models, we run a Monte Carlo simulation, randomly drawing 10,000 iterations from a Gaussian distribution of both our $T_{\text{eff}}$ ($\mu_{T_{\text{eff}}}$= 39,100K, $\sigma_{T_{\text{eff}}}=300$K) and $\log{g}$ ($\mu_{\log{g}}$= 8.69dex, $\sigma_{\log{g}}=0.04$dex) values.
For each realization, we perform a 2-D interpolation to determine the WD mass associated with that $T_{\text{eff}}$ and $\log{g}$ combination. This analysis returns a WD mass of $M_{\rm{WD}} = 1.06 \pm 0.02 M_{\odot}$ and a WD cooling age of $t_{\text{cool}}=21.87 \pm 4.85$ Myr. Figure \ref{fig:wdcooling} shows the results of the simulation for the thick WD model atmospheres for both $M_{\rm{WD}}$ and $t_{\text{cool}}$. Repeating the analysis with a thin hydrogen atmosphere model yields differences in both $t_{\rm cool}$ and $M_{\rm WD}$ that are smaller than the associated uncertainties. Figure \ref{fig:summarysystems} highlights Alessi12-PCE in the $P_{\rm orb}$–$M_{\rm WD}$ plane relative to the currently known population of detached WD+MS post-CE binaries.

The WD mass found here implies a current mass ratio for the Alessi12-PCE system of $q=$ $M_*$/$M_{\rm WD}$ $=0.35 \pm 0.04$ M$_\odot$. While this is smaller than the mass ratio implied by the ratio of the H$\alpha$ and WD semi-amplitudes measured in Section~\ref{sec:orbits}, we emphasize that because the H$\alpha$ emission in Alessi12-PCE feature likely originates on the irradiated surface of the M dwarf, it does not directly trace the center-of-mass of the M dwarf system (see Section~\ref{sec:reflection}). While we do not attempt to correct for this, as stated in Section \ref{sec:reflection}, previous work has found that the center-of-mass semi-amplitude in similar systems is larger than that measured from emission lines \citep[e.g.,][]{Parsons2010}. As such, the ratio of the semi-amplitudes provides an upper limit on the mass ratio of $q<0.5$, which is fully consistent with the mass measurements presented in this and the previous subsection.
Additional methods for measuring the WD mass in Alessi12-PCE will be presented in future work.

\subsection{Final Binary Separation}

From Keplers third law, the binary properties ($M_{*}$, $M_{\rm{WD}}$, and $P_{\text{orb}}$) of Alessi12-PCE imply a final separation of $a_{f}=2.070 \pm 0.023 R_{\odot}$. 

\section{Pre-CE Progenitor Properties}\label{sec:preCE}

To explore the full CE evolution for Alessi12-PCE, we reconstruct the pre-CE progenitor masses of both the M4V companion and the WD. We also discuss implications for the initial binary separation at the onset of CE evolution by using three fiducial interaction times during the AGB phase.

\subsection{M Dwarf Progenitor Mass}\label{sec:mdwarfprogen}
Low-mass M dwarfs ($\lesssim 0.35 M_{\odot}$) have almost fully convective interiors and as a consequence, have very long thermal timescales (generally $10^{8}-10^{9}$ years). In contrast, CE phases are very short, typically lasting on the order of only a few dynamical timescales  \citep[generally hundreds of days;][]{1996ApJ...471..366R, 2006MNRAS.370.2004N, 2012ApJ...746...74R, 2020MNRAS.495.4028C}. This difference in timescales is generally thought to prevent M dwarfs from accreting mass from the CE (or even the AGB star prior to the CE), allowing the mass and structure of M dwarfs to remain largely unchanged during a CE phase \citep{1991ApJ...370..709H, Nordhaus2013, 2018MNRAS.481.1083P}. As such, we assume that the pre-CE mass of the M4V star in Alessi12-PCE is, to first order, the same as its measured post-CE mass of $M_{*}=0.370 \pm 0.058 M_{\odot}$.

\subsection{White Dwarf Progenitor Mass}\label{sec:wdprogenitor}
Using our WD mass measurement, WD cooling age, and cluster age, we can determine Alessi12-PCE's WD progenitor mass, $M_{i}$. To simulate the pre-CE progenitor evolution, we use the Modules for Experiments in Stellar Astrophysics \citep[\texttt{MESA}; ][]{2011ApJS..192....3P} models outlined in \cite{WilsonNordhaus2019}. For progenitors of $M_{i}=1-6 M_{\odot}$, these models utilize \texttt{MESA} release 10108 and assume (i) solar metallically (consistent with the Alessi 12 open cluster) and (ii) mass-loss parameters/schemes for red giant branch (RGB) stars outlined in \cite{1977A&A....61..217R} ($\eta=0.7$) and AGB stars outlined in \cite{1995A&A...297..727B} ($\eta=0.15$). Importantly, these model parameters produce WDs that match the WD IFMR in \cite{Cummings2018} to within the measured errors, making them highly useful in investigating CE evolution \citep{WilsonNordhaus2019, 2021MNRAS.502L.110C, 2022MNRAS.516.2189W, 2022AJ....163...34M, 2022MNRAS.511.5994G, 2024arXiv240604118N, 2025PASA...42...27C}.

To constrain the pre-CE conditions of the Alessi12-PCE system, we assume that the onset of the CE phase coincides with the formation of the WD, since further WD growth would be truncated thereafter.  Following the standard approach used to derive the IFMR from cluster WDs \citep[e.g.,][]{Cummings2018}, the formation time is given by $t_{\text{form}} = t_{\text{cluster}} - t_{\text{cool}}$, where $t_{\text{cluster}}$ is the cluster age and $t_{\text{cool}}$ is the WD cooling age. The age of Alessi 12 is reported as $t_{\text{cluster}}=134.89$ Myr by \cite{CG2020}. Although individual cluster age uncertainties are not provided, \cite{CG2020} quote a typical uncertainty of $\log{t_{\text{cluster}}}=0.15-0.25$ for their catalog. We adopt an uncertainty of $\log{t_{\text{cluster}}}=0.20$, which corresponds to $t_{\text{cluster}}=134.89 \pm 49.78$ Myr. From Section \ref{sec:wdmasscooling}, we obtain $t_{\text{cool}}=21.87 \pm 4.85$ Myr. Combining these values yields a CE onset time of $t_{\text{form}} = 113.04 \pm 50.02$ Myr after cluster formation.

\begin{figure*}
    \centering
    \includegraphics[width=1\linewidth]{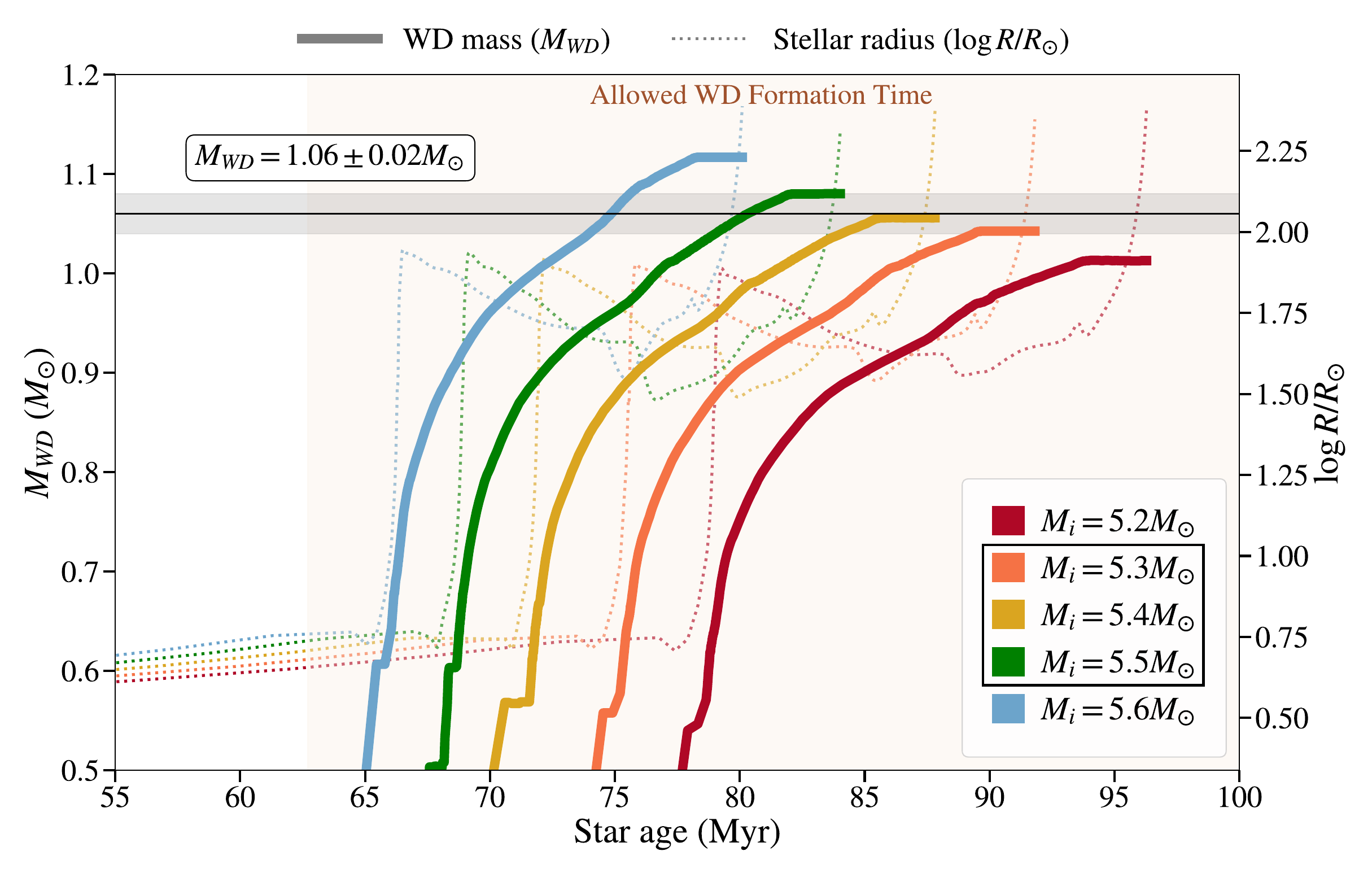}
    \caption{The pre-CE WD progenitor core mass (solid lines) and radius (dashed lines) evolution of $5.2-5.6 M_{\odot}$ \texttt{MESA} models from \cite{2020MNRAS.497.1895W}. Only models that (i) reach a WD mass of $M_{\text{WD}}=1.06 \pm 0.02 M_{\odot}$ (black line) in the time span of $t_{\text{form}} = t_{\text{cluster}} - t_{\text{cool}} = 113.13\pm51.29$ Myr and (ii) enter a CE phase on the AGB as opposed to the RGB are viable progenitor models (orange, yellow, and green lines). As such, we determine the WD progenitor mass is in the range of $M_{i}=5.3-5.5 M_{\odot}$ (boxed in legend) and thus adopt a pre-CE progenitor mass of $M_{i}=5.4 \pm 0.1 M_{\odot}$. 
    }
    \label{fig:mesamodels}
\end{figure*}

For each \texttt{MESA} model with initial masses $M_{i}=5.2, 5.3, 5.4, 5.5$ and $5.6 M_{\odot}$, we plot the helium core mass (solid lines), which corresponds to the eventual WD, and the stellar radius (dashed lines) as functions of stellar age in Figure \ref{fig:mesamodels}. 
It is worth noting that in these \texttt{MESA} models, the helium core mass reaches a plateau in its evolution during the AGB phase. However at much later time steps, a decline in the helium core mass is observed. The physical origin of the drop is uncertain, and we therefore adopt the value at the end of the plateau as the representative endpoint of the helium core's evolution in this study. In particular, detailed studies have found that the expected core growth on the AGB phase for stars in the mass range relevant for Alessi12-PCE are minimal: $\sim3.26-3.55 \times 10^{-7} M_{\odot}/\text{yr}$ for a 5.0-5.5 $M_{\odot}$ progenitor at solar metallicity \citep{2015MNRAS.446.2599D}. From our \texttt{MESA} models, we estimate an AGB lifetime of $\approx 0.1$Myr for our progenitor stars, which combined with the core growth rate above, would correspond to $\approx 0.05 M_{\odot}$ in growth over the full AGB phase. The shaded region in Figure~\ref{fig:mesamodels} indicates the allowed range for the onset of CE phase derived above. Despite the relatively large uncertainty in the Alessi 12 cluster age, the inferred formation times are inconsistent with a merger scenario (see Section \ref{sec:mergerproduct} for details), so the WD evolution can be modeled as a single star up until the CE event.

Within this framework, we find that only models with $M_{i}> 5.3 M_{\odot}$ produce helium core masses consistent with the post-CE WD mass ($1.06 \pm 0.02 M_{\odot}$) inferred for Alessi12-PCE. The large WD mass further implies that the progenitor evolved to the AGB prior to the onset of the CE phase, otherwise the core growth would have been truncated at an earlier evolutionary stage. We therefore require that, when $M_{\rm core}=1.06 \pm 0.02 M_{\odot}$ for a given model, the stellar radius exceeds the maximum radius attained during the RGB, ensuring that binary interaction did not occur on the RGB. This condition is not satisfied for progenitors with $M_{i} > 5.5 M_{\odot}$ (i.e., when their core masses are within the range found for Alessi12-PCE the radius of the progenitor is smaller than the maximum RGB radius). These requirements thus constrain the progenitor mass to $5.3 M_{\odot} < M_{i} < 5.5 M_{\odot}$, and we adopt $M_{i}=5.4 \pm 0.1 M_{\odot}$ as the fiducial initial mass. This yields initial mass ratios between $q=0.067-0.069$ for Alessi12-PCE.

\subsection{Initial Binary Separation} \label{sec:sep}

To estimate the initial binary separation at the onset of Roche lobe overflow ($a_{i}$), we assume that the stellar radius at the commencement of the CE phase ($R_{i}$) is equal to the Roche lobe radius ($R_{L}$). We compute $a_{i}$ using the \cite{1983ApJ...268..368E} approximation for the Roche lobe radius:

\begin{equation}
    \frac{R_{L}}{a_{i}} = \frac{0.49q^{2/3}}{0.6q^{2/3} + \ln(1 + q^{1/3})}
\end{equation}

where $q \equiv M_{*}/M_{i}$ is the mass ratio, $M_{*}$ is the mass of the M dwarf companion, and $M_{i}$ is the mass of the progenitor at the onset of Roche lobe overflow. From Figure~\ref{fig:mesamodels}, we see that multiple time steps along the AGB phase satisfy our constraints on the WD mass. This reflects a degeneracy not in the evolutionary stage itself, but in when along the AGB the CE event occurs, since the core mass is not expected to grow significantly during this phase \citep{2015MNRAS.446.2599D}. To capture this uncertainty, we evaluate three fiducial cases representing early, mid, and late AGB stages for models with $M_{i}=5.3,5.4$, and $5.5 M_{\odot}$. Specifically, we define the early AGB phase when the progenitor has reached 33\% of its maximum radius, the mid-AGB phase at 66\%, and the late AGB phase at 99\% of its maximum radius. We find  initial separations spanning $a_{i}\approx 1200-3400\, R_{\odot}$ depending on the assumed timing during the AGB phase. These values are subsequently used to estimate the CE efficiency parameters $\alpha_{\text{CE}}$ in a model agnostic way in Section \ref{sec:alphaest}. Table \ref{tab:fitderived} summarizes the derived post-CE and reconstructed pre-CE parameters for Alessi12-PCE.

\begin{table*}[ht]
\centering
\captionsetup{justification=centering}
\caption{Post-CE and pre-CE parameters for Alessi12-PCE.}
\label{tab:fitderived}

\renewcommand{\arraystretch}{1.15}
\setlength{\tabcolsep}{2pt} 
\begin{threeparttable} 
\begin{tabular}{lcl}
\hline
\multicolumn{3}{c}{\textbf{Post-CE System Parameters}} \\
\hline
\hline
Parameter & Value &  Notes/Units \\
\hline
 White dwarf type & DA &  White dwarf spectral type \\

$T_{\text{eff}}$ & $39,100 \pm 300$  & WD effective temperature [K] \\
 $\log g$ & $8.69 \pm 0.04$ & WD surface gravity $\left[\log_{10}\frac{g}{\mathrm{cm\,s^{-2}}}\right]$ \\
 $t_{\rm cool}$ & $21.87 \pm 4.85$  &  WD cooling age [Myr] \\
 $M_{\rm WD}$ & $1.06 \pm 0.02$  &  WD mass [$M_\odot$]\\
 M dwarf type & M4V   & M dwarf spectral type \\
 $M_{*}$\tnote{a} & $0.37 \pm 0.058$  &  M dwarf mass [$M_\odot$] \\
 $P_{\text{orb}}$ \phantom{X} & $6.99000 \pm 0.00001$ & Orbital period, final [hours] \\
 $a_f$ & $2.07 \pm 0.02$ & Orbital semi-major axis, final [$R_{\odot}$] \\ 
$q_{f}$ & $0.35 \pm 0.04$  & Final mass ratio $\left( \frac{M_{MS}}{M_{\rm{WD}}} \right)$ \\
\hline
\multicolumn{3}{c}{\textbf{Pre-CE System Parameters}} \\
\hline
\hline
Parameter & Value  & Notes/Units \\
\hline
 $M_i$ & $5.4 \pm 0.1$ & WD progenitor ZAMS mass [$M_\odot$] \\
$a_i$\tnote{b}  & $1200-3400$  & Orbital semi-major axis, initial $[R_{\odot}]$  \\
$q_{i}$ & $0.070 \pm 0.0084$  & Initial mass ratio $\left( \frac{M_{MS}}{M_{i}} \right)$ \\
\end{tabular}

\begin{tablenotes}
\footnotesize
\item[a] We assume that the M dwarf’s post-CE mass ($M_{*}$) is equal to its zero-age main sequence (ZAMS) progenitor mass, since the thermal timescale of an M dwarf ($10^{8}-10^{9}$ years) is much longer than the duration of the CE phase (hundreds of years). This allows the M dwarf’s mass and internal structure to remain effectively unchanged during the CE (see Section \ref{sec:mdwarfprogen}).
\item[b] Range reflects three assumed CE interaction stages along the AGB for progenitor 
masses $M_i = 5.3$--$5.5\,M_\odot$; see Section~\ref{sec:sep}. 
\end{tablenotes}
\end{threeparttable}

\end{table*}

\section{CE Physics with Alessi12-PCE} \label{sec:discussion}

The CE ejection efficiency parameter, $\alpha_{\text{CE}}$,
describes how efficiently the orbital energy released during inspiral ($\Delta E_{\text{orb}}$) is used to unbind the envelope ($E_{\text{bind}}$) during a CE interaction \citep{Paczynski1976, Webbink1984}: $\alpha_{\text{CE}} = {E_{\text{bind}}} / {\Delta E_{\text{orb}}}$. Below, we present our own empirical (model agnostic) determination of the range of $\alpha_{\text{CE}}$ parameters that are consistent with Alessi12-PCE. We then show how the post-CE parameters (i.e., $a_{f}$) of Alessi12-PCE can be reproduced using convective CE models and place this work in the context of previous efforts to constrain $\alpha_{\text{CE}}$ with WD binaries.

\subsection{Model Agnostic $\alpha_{\text{CE}}$ Estimates Reveal the Impact of CE Onset Time} \label{sec:alphaest}

To determine the $\alpha_{\text{CE}}$ required to reproduce the post-CE conditions of Alessi12-PCE, we evaluate three representative stages along the AGB (early, mid, and late times), which we define as a fraction of the maximum radial extent of the AGB star (33\%, 66\%, and 99\%) for each allowed $M_{i}$ (see Section \ref{sec:sep}). This approach accounts for the fact that, while the progenitor clearly reached the AGB before entering the CE (given its high mass), the precise timing of CE onset along the AGB is uncertain. For each progenitor and AGB stage, we compute $E_{\text{bind}}$ directly from the \texttt{MESA} stellar structure model at the corresponding timestep. The initial orbital energy, $E_{\text{orb},i}$, is calculated using the progenitor and companion masses together with the initial separation, $a_{i}$, while the final orbital energy, $E_{\text{orb},f}$, is derived from the measured post-CE WD and M4V companion masses and the observed orbital separation. 

Our results indicate that the inferred $\alpha_{\text{CE}}$ depends strongly on when the CE interaction occurs along the AGB: near the tip (99\% of the maximum AGB radius), $\alpha_{\text{CE}} \approx 0.034-0.058$; at mid-AGB (66\% of the maximum AGB radius), $\alpha_{\text{CE}} \approx 0.35-0.65$; and for an earlier interaction (33\% of the maximum AGB radius), the inferred efficiency rises to values as high as $\alpha_{\text{CE}} \approx 1.5$. Hence, $\alpha_{\text{CE}}$ values for Alessi12-PCE can vary widely, even though the WD helium core mass grows very little on the AGB. This is because the progenitor radius at the onset of the CE phase strongly affects the envelope binding energy. Larger radii at later AGB stages correspond to lower binding energies, making envelope ejection easier and resulting in smaller inferred $\alpha_{\text{CE}}$ values to reached a fixed final separation.  In contrast, CE interactions that occur earlier on the AGB (with smaller radii) require higher $\alpha_{\text{CE}}$ values to unbind the envelope. We note that efficiency values exceeding $\sim$1 generally indicate that additional energy contributions may have been required to unbind the envelope \citep{2007MNRAS.376..599N,2011PNAS..108.3135N,Ivanova2015,2018MNRAS.480.1898C,2019MNRAS.490.1179G,2022MNRAS.511.5994G,2024MNRAS.528..234C}. However more broadly, this example highlights how sensitive $\alpha_{\text{CE}}$ is to the progenitor radius at CE onset, where even modest changes in progenitor radius can produce substantial differences in the inferred $\alpha_{\text{CE}}$. Without an independent constraint on when along the AGB the CE interaction occurs, $\alpha_{\text{CE}}$ cannot uniquely be determined, and instead we must infer a range of possible values that reflect this uncertainty. 

\subsection{Convective CE Models Reproduce Alessi12-PCE}\label{sec:convection}

Here, we compare the parameters of the Alessi12-PCE system to those predicted by a the physically-motivated theoretical CE framework presented in \cite{WilsonNordhaus2019,2022MNRAS.516.2189W}, which models the inspiral of a companion during the CE taking into account potential energy loses due to the presence of a surface convection zone. Our goal is to assess whether there is a specific time for the onset of the CE phase on the AGB where the observed final separation ($a_{f}$) of the Alessi12-PCE system is reproduced.

\subsubsection{Summary of Convective CE Models}

The key physical concepts behind the model of \cite{WilsonNordhaus2019} can be summarized as: 
\begin{enumerate}
    \vspace{-0.1in}
    \item During a CE event, as the orbit decays, the energy released will work to unbind the envelope \emph{unless} it is lost at the stellar surface via radiation.
    \vspace{-0.1in}
    \item Post-MS stars have large and vigorous convective envelopes that distribute energy homogeneously and transport orbital energy to the surface of the star.  If convective transport to the surface is faster than the orbit decays, energy is lost from the system via radiation and the CE self-regulates.  If the orbit decays faster than convection can transport energy to the surface, the energy cannot escape and must contribute toward envelope unbinding.
    \vspace{-0.1in}
    \item Taken together, points 1.\ and 2.\ imply that for certain stellar structures, there may be regions of the primary star where convection is able to carry orbital energy liberated during a CE event to the surface---thereby lowering the overall CE efficiency.
    \vspace{-0.1in}
\end{enumerate}

As such, \cite{WilsonNordhaus2019} defined a physically-motivated framework to model the inspiral of a companion into a primary star stellar profile while testing where convection will impact the CE efficiency and predicting the final outcome of the CE event. Specifically, at each timestep of the inspiral they check whether (i) the companion is located within a surface contact convection zone of the primary, (ii) the convective timescale is shorter than the inspiral timescale, and (iii) the drag luminosity generated by the inspiral is less than the maximum additional luminosity that can be carried by convection. If all conditions are met, they assume that the CE efficiency \emph{for that timestep} is zero---i.e., that \emph{none} of the change in orbital energy generated during that timestep goes towards unbinding the envelope (instead it is carried to the surface and radiated away). In contrast, if the companion is located within a region of the primary star that is not a surface contact convection zone, or where the inspiral timescale is shorter than the convective timescale\footnote{In practice, \cite{WilsonNordhaus2019} found that the drag luminosity generated was almost always less than the maximum that could be transported via convection.} then the CE efficiency \emph{for that timestep} is assumed to be one--i.e., that \emph{all} of the change in orbital energy from that timestep goes towards unbinding the envelope.

With these assumptions, the model of \cite{WilsonNordhaus2019} can then estimate (i) the final orbital separation of a given CE as the point at which the total energy injected into the envelope exceeds the binding energy of the envelope above that point and (ii) a final global CE efficiency defined as the fraction of the total orbital energy released that went towards unbinding the envelope.  This model has been applied to the following three observed post-CE populations with remarkable success:  

\vspace{-0.1in}
\begin{enumerate}
    \item \emph{WD+M dwarf post-CE binaries:} The vast majority of observed WD+M dwarf post-CE binaries have sub-day periods, a fact that population synthesis studies have struggled to reproduce \citep{Toonen2017}.  However, this observational feature naturally emerges when comparing the orbital decay timescale to the convective transport timescale. For these systems, convection is fast and the orbit decays until it reaches sub-day orbital periods, at which point the energy transport to the surface is too slow and the energy is sufficient to unbind the envelope \citep{WilsonNordhaus2019}.
    \vspace{-0.1in}
    \item  \emph{Close double WD binaries:} Double WD systems are another example of post-CE binaries where population synthesis studies do not reproduce the observed distribution when using constant CE efficiencies \citep{Toonen2017}.  However, when the effects of convective transport and radiative losses are incorporated in CEs with this framework, the predicted post-CE separations closely track the observed double WD orbital parameter space \citep{2020MNRAS.497.1895W}.
    \vspace{-0.1in}
    \item \emph{High-mass post-CE stellar binaries:} While observations of low-mass post-CE systems show that the CE phase must be highly inefficient, observations of post-CE high-mass stellar binaries show that the CE phase must be highly \emph{efficient} \citep{2022MNRAS.516.2189W}.  Despite robust and vigorous convection in high mass stars, this is naturally explained as the orbit decays well before convection can transport the orbital energy to the surface. 
    \vspace{-0.1in}
\end{enumerate}
 
 In summary, the inclusion of convection in CEs reproduces observations of low-mass and high-mass binaries, is physically motivated, and remains a necessary ingredient for determining outcomes of CE evolution. 

\begin{figure*}
    \centering
    \includegraphics[width=1\linewidth]{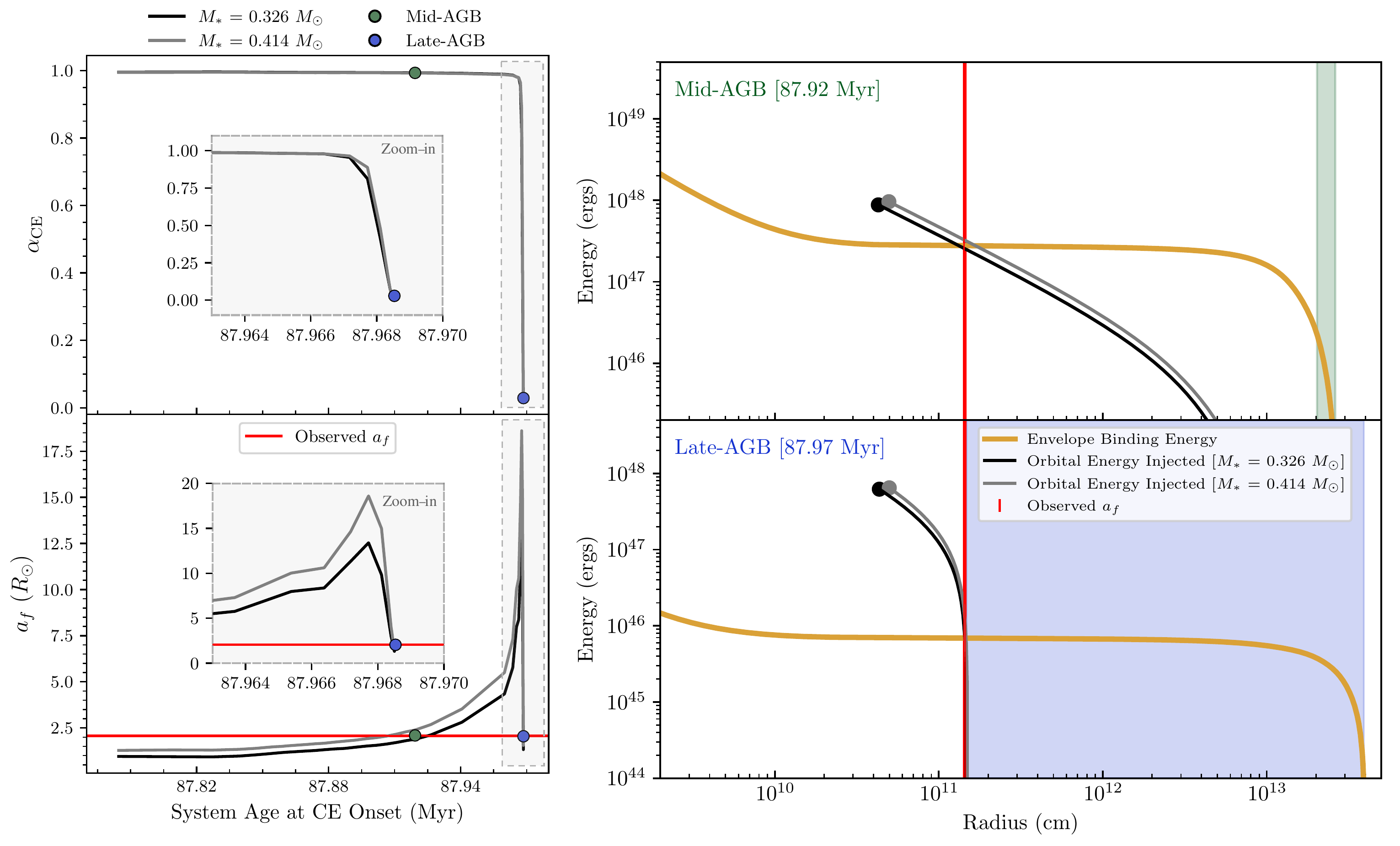}
    \caption{\textit{Left panels:} The CE ejection efficiency $\alpha_{\text{CE}}$ (top) and the final orbital separation (bottom) as a function of age along the AGB for a 5.4$M_{\odot}$ progenitor. The mid-AGB (green) and late-AGB (blue) \texttt{MESA} profiles from \cite{WilsonNordhaus2019} that reproduce the observed final separation of Alessi12-PCE (red line) are marked as circles. The two AGB stages for which envelope ejection occurs at the observed orbital separation differ dramatically in their required CE efficiency: $\alpha_{\text{CE}} \approx 0.99$ and $\alpha_{\text{CE}} \approx 0.05$. Insets show zoomed-in views of the shaded regions, highlighting the rapid evolution near these solutions. \textit{Right panels:} The binding and orbital decay energy profiles of the same two exact AGB evolutionary stages that can reproduce the final separation of Alessi12-PCE. The mid-AGB (top; green) and late-AGB (bottom; blue) plots depict where inside the CE companions of $M_{*} = 0.326 M_{\odot}$ and $M_{*} = 0.414 M_{\odot}$ eject the envelope. Shaded blue and green regions represent the portion of the primary’s radius where rapid convection efficiently transports orbital energy to the surface, where it is radiated away. The corresponding black and gray circles show where within the CE those same companions would tidally disrupt if inspiral had continued.}
    \label{fig:ConvectiveCEModels}
\end{figure*}

\subsubsection{Application of Convective CEs to Alessi12-PCE}

When applying this framework to the Alessi12-PCE system, we select a set of $\sim$30 stellar profiles from the \texttt{MESA} models of each of the 5.3, 5.4, and 5.5 M$_\odot$ primary stars described above. By default, we analyze every 50 \texttt{MESA} timesteps during the AGB phase until the point of maximal radial extent. Analysis on a denser time sampling is then performed in regions where the final predicted CE outcome more closely matches that of the Alessi12-PCE system (see below). For each stellar profile considered, we model the inspiral of both a $M_{*}$ = 0.326 M$_\odot$ and $M_{*}$ = 0.414 M$_\odot$ companion star (which span the range possible companion masses for Alessi12 found in Section~\ref{sec:mdwarfproperties}) and determine the the final separation and CE efficiency predicted. 

The results of this analysis for the 5.4 M$_\odot$ primary star are shown in Figure~\ref{fig:ConvectiveCEModels}. On the left, we show the predicted evolution of the final post-CE binary separation and global CE efficiency as a function of time, as the primary star evolves up the AGB. For much of the AGB phase, the inspiral timescale for $M_{*}$ = 0.370 $\pm$ 0.058 M$_\odot$ companions is shorter than the convective timescales in all but the outer $\lesssim$20\% of the envelope. As a result, convection cannot effectively carry away newly liberated orbital energy and the predicted CE efficiencies are nearly unity. During these phases (when the CE efficiency is high) the predicted final separation of the binary gradually increases as the star evolves up the AGB. This is because the binding energy of the envelope decreases as the radius expands, so less orbital energy is required to eject it. We find that the predicted final separation closely matches that observed for the Alessi12-PCE system when the radius of 5.4 M$_\odot$ primary star is approximately 360--390 R$_\odot$ ($\sim$50-55\% of its maximum extent). This occurs when the star is $\sim87.93$ Myr old. 

However, as the primary star continues to evolve up the AGB and the envelope becomes sparser, the inspiral timescale for the companion star will increase\footnote{See \cite{WilsonNordhaus2019} equation 3; the inspiral timescale scales with the inverse of the stellar density.}. Eventually, the inspiral timescale for a $M_{*}$ = 0.370 $\pm$ 0.058 M$_\odot$ companion increases enough that it exceeds the convective timescales in a much larger fraction ($\gtrsim 99\%$) of the envelope. This change results in a (precipitous) drop in predicted CE efficiency, as convection can effectively carry away some of the liberated orbital energy. This drop in CE efficiency is accompanied by a decrease in the predicted final post-CE binary separation, as the companion must inspiral further before the change in orbital energy contributes to ejecting the envelope. As a result, we find that the predicted final separation \emph{again} closely matches that observed for the Alessi12-PCE system when the 5.4 M$_\odot$ primary star is approximately 625--630 R$_\odot$ ($\gtrsim$99\% of its maximal extent). This occurs when the star is $\sim$87.97 Myr old. Both the 5.3 and 5.5 M$_\odot$ models show qualitatively similar behavior to the 5.4 M$_\odot$ model when analyzed in the framework of \cite{WilsonNordhaus2019}.

\begin{figure*}
    \centering
    \includegraphics[width=1\linewidth]{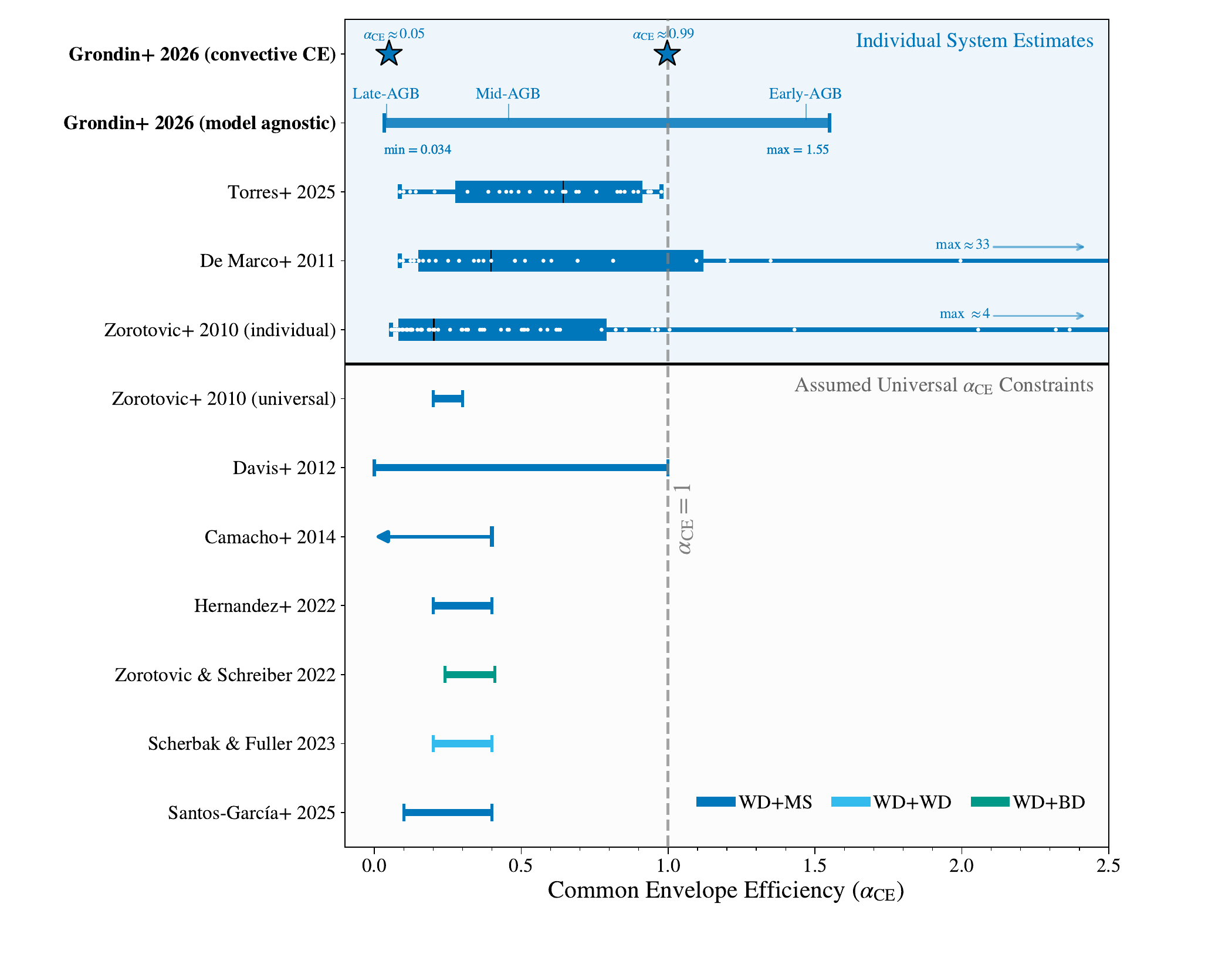}
    \caption{Comparison of CE efficiency ($\alpha_{\text{CE}}$) estimates for WD post-CE binaries from this work and the literature, color-coded by binary type (WD+MS, WD+WD, WD+brown dwarf). The top panel shows literature which present individual system estimates for $\alpha_{\text{CE}}$. This includes our model agnostic range of $\alpha_{\text{CE}}\approx 0.03-1.55$, which includes the mean measurements for the early, mid, and late AGB CE onset times across our three progenitor masses, along with the $\alpha_{\text{CE}}$ values for the two distinct evolutionary times where convective CE models reproduce the final orbital separation of Alessi12-PCE ($\alpha_{\text{CE}}\approx 0.05$ and $\approx 0.99$; starred). For comparison, we show the individual system estimates of \citet{Zorotovic2010}, \citet{2011MNRAS.411.2277D}, and \citet{2025A&A...698A.173T}, with individual measurements over-plotted as points on each boxplot. The bottom panel shows constraints derived under the assumption that a single, universal $\alpha_{\text{CE}}$ value represents the entire population \citep{Zorotovic2010, 2012MNRAS.419..287D, Camacho2014, 2022MNRAS.513.3587Z, 2022MNRAS.512.1843H, Scherbak2023, 2025A&A...695A.161S}. These $\alpha_{\text{CE}}$ estimates typically cluster around $\alpha_{\text{CE}} \approx 0.2-0.4$, which reflects both the imposed assumption of $\alpha_{\text{CE}}$ being universal and, in several cases, an explicit $\alpha_{\text{CE}} < 1$ prior. We note that this figure only includes studies which measure the $\alpha_{\text{CE}}$ parameter directly, but recognize many additional efforts to constrain the $\alpha\lambda$ prescription have been made \citep[e.g.,][]{Nelemans2000, Toonen2013}.}
    \label{fig:alphacomparison}
\end{figure*}

Thus, we find that a convective CE model is able to reproduce the properties of the Alessi12-PCE system \emph{either} at roughly the mid-point of the AGB (with a high CE efficiency of $\alpha_{\text{CE}} \gtrsim 0.99$) or near the end of the AGB (with a low CE efficiency of $\alpha_{\text{CE}} \lesssim 0.05$). We highlight examples of both of these cases in the righthand panels of Figure~\ref{fig:ConvectiveCEModels} and plot the $\alpha_{\text{CE}}$ measurements that reproduce Alessi12-PCE relative to the broader literature at both time-steps in Figure \ref{fig:alphacomparison} (see Section~\ref{sec:alphaconstraints} for further discussion of this comparison). In the righthand panels of Figure \ref{fig:ConvectiveCEModels}, the gold line shows the cumulative binding energy of the envelope above a given radius while the grey/black lines show the total orbital energy released \emph{that goes towards unbinding the envelope} as the companion inspirals. The point at which these two lines cross indicates where the envelope will be ejected. For both of these examples shown, this point matches the measured post-CE separation for the Aless12-PCE system (vertical red line). Also shown in each panel (green/blue shaded regions) are the portions of the envelope where convection can effectively carry the energy released during orbital decay to the surface where it is lost as radiation. For the top panel (mid-AGB) this is only a small portion of the outer envelope and a gradual increase in the amount of orbital energy injected into the envelope is then observed. In contrast, for the lower panel (late-AGB) this region extends significantly further into the AGB envelope and once the companion exits this region the change in orbital energy quickly surpasses the amount required to unbind the (sparse) envelope. 

Together these examples highlight that models in which convection sets the CE efficiency can reproduce the observed properties of the Alessi12-PCE system. However, consistent with what was found in Section~\ref{sec:alphaest}, the predicted $\alpha_{\text{CE}}$ can vary dramatically depending on the initial separation of the system and thus \emph{when} during the AGB phase the CE phase commences. Notably, if the CE phase of the Alessi12-PCE had been observed by a modern time-domain survey, it would have been possible to distinguish between the two scenarios highlighted in Figure~\ref{fig:ConvectiveCEModels}. In particular \cite{2024arXiv240604118N} find that in cases where convection carries significant orbital energy to the surface (as is the case for the late-AGB solution described above) the timescale for the observed light curve will lengthen dramatically, showing a slow rise over years before a short-lived plateau phase. Surveys such as ZTF and the Vera C. Rubin Legacy Survey of Space and Time (LSST) could therefore further elucidate the role of convection in CE events.

\subsection{Alessi12-PCE Relative to Previous $\alpha_{\text{CE}}$ Estimates} \label{sec:alphaconstraints}

There is a rich history of studies that attempt to constrain $\alpha_{\text{CE}}$ with WD post-CE binaries \citep[e.g.,][]{2001A&A...365..491N, Zorotovic2010, 2011MNRAS.411.2277D, 2012MNRAS.419..287D, Toonen2013, Camacho2014, 2021MNRAS.501.1677H, 2022MNRAS.517.2867H, 2022MNRAS.512.1843H, 2022MNRAS.513.3587Z, Scherbak2023, 2025A&A...695A.161S, 2025A&A...698A.173T}. Many of these studies reconstruct the pre- and post-CE properties of individual binaries and derive a corresponding value of $\alpha_{\text{CE}}$ for each system, but then use the full population to constrain a single, universal value of $\alpha_{\text{CE}}$. We compare a number of these literature estimates with our own model agnostic $\alpha_{\text{CE}}$ constraints for Alessi12-PCE from Section~\ref{sec:alphaest}, as well as the two convective CE solutions identified in Section~\ref{sec:convection}, in Figure \ref{fig:alphacomparison}.

From Figure \ref{fig:alphacomparison}, literature $\alpha_{\text{CE}}$ estimates for WD post-CE binaries are generally less than 1, clustering around $\alpha_{\text{CE}} \approx 0.2-0.4$. Initially, it appears that our $\alpha_{\text{CE}}$ estimates are in tension with this previous work: our model agnostic values span $\alpha_{\text{CE}}\approx 0-1.5$, and our convective CE models lie at the extremes of the typical allowed range ($\alpha_{\text{CE}} \approx 0.05$ and 0.99). However, before combining individual measurements into a single estimate, several of these previous studies \citep[e.g.,][]{Zorotovic2010, 2022MNRAS.513.3587Z, Scherbak2023} report a large allowed $\alpha_{\text{CE}}$ range on a system-by-system basis, often comparable to or even exceeding the model-agnostic $\alpha_{\text{CE}}$ range for Alessi12-PCE. Moreover, many of these works restrict their analysis to $\alpha_{\text{CE}} < 1$ \citep[e.g.,][]{ 2021MNRAS.501.1677H, 2022MNRAS.517.2867H, 2022MNRAS.512.1843H, 2022MNRAS.513.3587Z}, which, by construction, excludes values beyond this range. Whether  $\alpha_{\text{CE}} > 1$ is physically viable remains contented, however we emphasize that many of these individual post-CE binaries could potentially be explained by a wider spread in $\alpha_{\text{CE}}$. 

This is particularly clear from a handful of studies that do report $\alpha_{\text{CE}}$ on a system-by-system basis, where these binaries can take on a wide array of $\alpha_{\text{CE}}$ values 
and are scattered across much of the same range we infer with Alessi12-PCE. For example, although \cite{Zorotovic2010} conclude that $\alpha_{\text{CE}} \approx 0.2-0.3$ is the most plausible global value across their entire sample, when reconstructing the evolution of $\approx 75$ WD+MS post-CE binaries identified via the Sloan Digital Sky Survey (by incorporating the internal energy of the envelope into the binding energy parameter $\lambda$ rather than assuming a fixed value), their individual system estimates span $\alpha_{\text{CE}} \approx 0.05-4.10$. Using an independent reconstruction of progenitor masses and radii via WD IFMRs and stellar evolution models, \cite{2011MNRAS.411.2277D} find an even broader range for a sample of $\approx 30$ binaries compiled from \cite{2003A&A...406..305S}, \cite{2009PASP..121..316D}, and \cite{Zorotovic2010}: $\alpha_{\text{CE}} \approx 0.09-33$. Most recently, \cite{2025A&A...698A.173T} present an inverse binary population synthesis approach (using the \texttt{MRBIN} code), where their model populations are tuned to reproduce the properties of $\approx$ 30 eclipsing post-CE binaries from \citet{2023MNRAS.521.1880B}. From this analysis, \cite{2025A&A...698A.173T} recover another wide range of $\alpha_{\text{CE}} \approx 0.09-0.98$. Together, these individual system estimates span values well beyond the widely-used $\alpha_{\text{CE}} \approx 0.2-0.4$ range commonly explored in binary population synthesis studies (in some cases, by more than an order of magnitude). In combination with our presented $\alpha_{\text{CE}}$ estimates of Alessi12-PCE, this underscores that a single, universal value of $\alpha_{\text{CE}}$ -- or even a small, restricted range -- is unlikely to capture the true diversity of CE outcomes in WD post-CE binaries. 

\section{Testing Alternative Formation Pathways of Alessi12-PCE} \label{sec:alternate}

In order to use Alessi12-PCE as a benchmark to examine CE evolution, it is essential to determine if the current WD in the system can be modeled using single star evolution prior to CE evolution, or if binary interactions altered the evolution prior to the CE event. Here, we consider two alternative formation scenarios: (i) a binary merger creating a massive WD and (ii) a dynamical formation channel.

\subsection{Merger Product?}\label{sec:mergerproduct}

The average mass of a WD which arises from single star evolution is $M_{\rm{WD}}\approx0.6 M_{\odot}$. High-mass ($\approx 1.0 M_{\odot}$) WDs are rare, comprising only $\approx 5-10\%$ of the WD population \citep[for a recent observational review, see][]{2025arXiv250219496P}. Importantly, a significant fraction of massive WDs are expected to originate from binary mergers rather than from single-star evolution, with merger fractions of $30-45\%$ estimated for WDs having $M_{\rm{WD}} > 0.9 M_{\odot}$ within 100 pc of the Sun \citep{2020A&A...636A..31T, 2023MNRAS.518.2341K}. In Section \ref{sec:wdmasscooling}, we measure the current WD mass of Alessi12-PCE to be $M_{\rm{WD}}=1.06 \pm 0.02 M_{\odot}$. Since this WD is massive, we consider the possibility that Alessi12-PCE is the result of a merger product (as in the case of the V471 Tau WD+MS post-CE binary; \citealt{2022AJ....163...34M}). 

\begin{figure}
    \centering
    \includegraphics[width=1.0\linewidth]{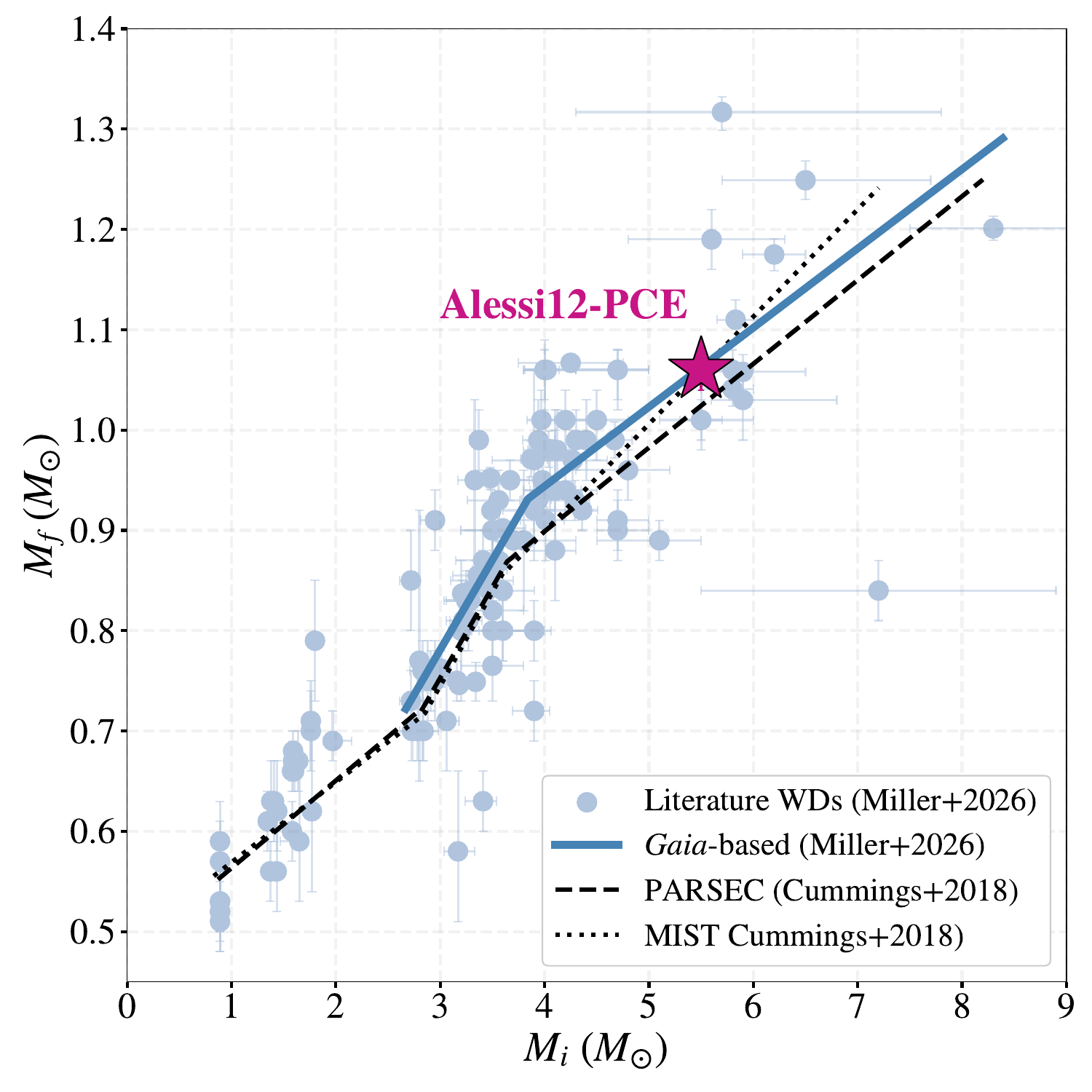}
    \caption{Alessi12-PCE (pink) is shown relative to the initial–final mass relation (IFMR) for single white dwarfs in open clusters from \cite{2026ApJ...996...69M} (blue). Empirical IFMR constraints from \cite{2026ApJ...996...69M} (blue line), along with relations based on PARSEC and MIST stellar models from \cite{Cummings2018} (black lines), are included for comparison. The inferred initial and final masses of Alessi12-PCE are consistent with the IFMR, suggesting that despite its high $M_{\mathrm{WD}}$, it likely formed through single-star evolution rather than a merger.}
    \label{fig:wdifmr}
\end{figure}

First, in Section~\ref{sec:wdprogenitor}, we noted that the cluster age of Alessi 12 is consistent with systems that grow to reach such a WD mass through single star evolution (see Figure \ref{fig:mesamodels}).
In this case, we used \texttt{MESA} models to measure a WD progenitor mass of $M_{i} = 5.4 \pm 0.1 \,M_{\odot}$. In Figure \ref{fig:wdifmr}, we highlight Alessi12-PCE relative to the \cite{2026ApJ...996...69M} observationally-derived IFMR for single WDs in open clusters from \textit{Gaia} DR3. This work contains a full sample of 122 WDs with final masses between $M_{\rm{WD}} = 0.510-1.317 \,M_{\odot}$ and inferred initial progenitor masses between $M_{i} = 0.890-8.300 \,M_{\odot}$. We also plot the IFMR derived by \cite{Cummings2018} for comparison. The final and initial WD masses of Alessi12-PCE are consistent with both literature IFMRs, highlighting that the mass (and temperature) of the WD in Alessi12-PCE is consistent with single-star evolutionary models. 

\subsection{Dynamical Formation?}

The high number densities in the cores of star clusters (mainly, globular clusters) frequently result in the dynamical formation of an array of compact object binaries \citep[e.g.,][]{2006MNRAS.372.1043I, 2008MNRAS.386..553I, 2021ApJ...917...28K}. Although Alessi12-PCE is a member of an \textit{open} star cluster, we consider the possibility that it could have a dynamical origin. As a first check, we consider only the influence of single-single encounters \citep[where binary-single, binary-binary, and higher order encounters with triple systems would be much rarer;][]{2011MNRAS.410.2370L}. Below, we compute the rate of single-single encounters, $\tau_{1+1}$, that occur in a standard OC using an interaction rate equation from \cite{2011MNRAS.410.2370L}:

\begin{align}
\tau_{1+1} =\ & 1.1 \times 10^{11} (1 - f_{b} - f_{t})^{-2} \left(\frac{1\ \text{pc}}{r_{c}}\right)^{3} \left(\frac{10^{3}\ \text{pc}^{-3}}{n_{0}}\right)^{2} \nonumber \\
& \ \ \ \ \ \ \  \left(\frac{v_{\text{rms}}}{5\ \text{km}\ \text{s}^{-1}}\right) \left(\frac{0.5\ M_{\odot}}{\langle m \rangle}\right) \left(\frac{0.5\ R_{\odot}}{\langle R \rangle}\right) \ \text{years}
\end{align}

Here, $f_{b}$ and $f_{t}$ are the binary and triple fractions of the cluster, $r_{c}$ is the core radius, $n_{0}$ is the number density, $v_{rms}$ is the velocity dispersion and $\langle m \rangle$ and $\langle R \rangle$ are the average mass and radius of a star in the cluster. Unfortunately, the Alessi 12 open cluster is poorly studied, and lacks observational measurements of all of the required parameters to compute $\tau_{1+1}$. Hence, to estimate $\tau_{1+1}$ for Alessi12-PCE, we sample over a wide range of potential cluster properties.

Beginning with our age estimate of Alessi 12 from Section \ref{sec:wdprogenitor} of $134.89 \pm 49.78$ Myr, we randomly sample 10,000 cluster ages between 50 and 150 Myr. To generate possible cluster populations for each age, we sample stars from one of five evolved 10,000 $M_\odot$ stellar populations with ages of 50 Myr, 75 Myr, 100 Myr, 125 Myr, and 150 Myr using \texttt{PARSEC} version 1.2S\footnote{\href{http://stev.oapd.inaf.it/cmd}{http://stev.oapd.inaf.it/cmd}}\citep{Bressan2012, Chen14, Chen15, Tang14, Marigo17, Pastorelli19, Pastorelli20}. Here, we assume a \cite{Kroupa01} initial mass function, a distance modulus of 8.9, reddening of 0.08 \citep{Kharchenko05} and a solar metallicity. \cite{CG2020} observe 307 high-probability member stars with  \textit{Gaia} magnitudes brighter than 18 mag in Alessi 12, so we assume these observations are complete and stop sampling stars once we build a population of 307 stars. This exercise yields 10,000 possible cluster masses with a mean total mass estimate of $536.24 \pm 22.08 M_{\odot}$ for the Alessi 12 OC.

To estimate the density profile of Alessi 12, we use the \citet{CG2020} half-number radius of 0.682 degrees from the cluster's center. Given the cluster's young age, there has likely been very little mass segregation in Alessi 12. Hence, we assume the cluster's half-mass radius is also 0.682 degrees, but allow for an uncertainty of $10\%$ when sampling possible half-mass radii. With little known about the cluster's structural properties, we simply assume Alessi 12 can be represented by a lowered isothermal model as described in \cite{2015MNRAS.454..576G}. We sample 10,000 density profiles where the profile's shape is governed by the truncation parameter $g$ and the central density is governed by the parameter $\phi(r=0)$. As described in \cite{2015MNRAS.454..576G}, $g$ is sampled between 0.01 and 3.49 and $\phi(r=0)$ is sampled between 1 and 20. We accept any possible profile with a virial radius to tidal radius ratio greater than 0.64, which is consistent with realistic fits to star clusters \citep{2015MNRAS.454..576G, Wen2024}. 

For each of the 10,000 combinations of randomly sampled ages, cluster masses, sizes, and density profiles, we initially calculate $\tau_{1+1}$ by assuming a multiple star fraction ($f_b +f_t$) of $35\%$. The distribution of possible $\tau_{1+1}$ values is double peaked, as both $g$ and $\phi_0$ can yield high core densities. The peaks occur near $10^{13}$ and $10^{14.25}$ years, with a dispersion on the order of $10^{12}$ years. Out of all our iterations, the smallest calculated value of $\tau_{1+1}$ is $10^{11}$ years. Hence, it would take longer than a Hubble time for an Alessi 12-like cluster to undergo a single-single dynamical interaction that could form a binary star. An extremely high multiple fraction of $100\%$ \citep[which is unphysically high, as shown in][]{2023A&A...675A..89D} only decreases our estimate of $\tau_{1+1}$ by an order of magnitude. Conversely, focusing on density profiles where the ratio of the outermost star's distance from the center to the half-mass radius matches what is observed for Alessi 12 increases our estimated $\tau_{1+1}$ by an order of magnitude. While it is possible that the  \cite{2011MNRAS.410.2370L} present day $\tau_{1+1}$ could be different than $\tau_{1+1}$ at earlier points in Alessi 12's evolution, we conclude that it is extremely unlikely that Alessi12-PCE has a dynamical origin.

\section{Summary and Future Outlook} \label{sec:conclusions}

In this work, we present a new framework to infer the pre- and post-CE properties of WD+MS binaries using open star clusters. Here, we present the first demonstration of this method through the characterization of Alessi12-PCE--just the third detached WD+MS post-CE binary ever discovered in an open cluster, and the first for which both the pre- and post-CE properties are precisely measured. Our key conclusions are as follows:

\begin{enumerate}
    \item Modeling a Keck/LRIS spectrum yields an M4V companion ($M_{*}=0.37\pm0.058 \,M_{\odot}$) and a massive WD ($M_{\mathrm{WD}}=1.06\pm0.02 \,M_\odot$), while photometric (ZTF light curve) and spectroscopic (Gemini/GMOS and Shane/Kast radial velocity) observations establish a short orbital period of $P_{\text{orb}}\approx7$ hr, typical of post-CE WD+MS binaries.  Alessi12-PCE is the first post-CE binary in a stellar cluster found to host a WD with $M_{\mathrm{WD}} > 1\,M_\odot$. 
    
    \item Combined with the cluster age and comparison to the WD IFMR \citep{2026ApJ...996...69M}, we find that despite its high mass, the WD in Alessi12-PCE is consistent with single-star evolution rather than a merger in a binary. Moreover, we find that it would take longer than a Hubble time for an Alessi12-like cluster to undergo a dynamical single-single interaction that could form a binary star. This makes Alessi12-PCE an excellent benchmark to test CE physics.
   
    \item Using \texttt{MESA} models, we reconstruct the initial pre-CE progenitor mass to be $M_{i} = 5.4 \pm 0.1 M_{\odot}$. As just the first cluster WD+MS post-CE system with precise pre- and post-CE constraints, Alessi12-PCE enables direct tests of CE physics. By sampling CE onset for each progenitor mass across different stages during the AGB phase (e.g., early, mid and late-type AGB stars), we find a wide range of possible CE ejection efficiencies ($\alpha_{\mathrm{CE}} \approx \ $0.05–$1.35$) depending on the assumed time of CE onset. Interactions near the tip of the AGB (larger radii, lower binding energies) yield low efficiencies, whereas earlier AGB interactions (smaller radii, higher binding energies) require higher efficiencies. This demonstrates that the progenitor radius, and ultimately the timing of CE commencement, is degenerate with the inferred $\alpha_{\mathrm{CE}}$ itself, highlighting a key limitation of solely using heuristic prescriptions to understand CE evolution. 

\item  We find that the \cite{WilsonNordhaus2019} convective CE model is able to reproduce the observed properties of the Alessi12-PCE system at two times during the AGB evolution of Alessi12-PCE: at roughly the mid-point of the AGB (when it predicts a high CE efficiency of $\alpha_{\mathrm{CE}} \gtrsim 0.99$) or near the end of the AGB (when it predicts a low CE efficiency of $\alpha_{\mathrm{CE}} \lesssim 0.05$). This indicates that, for this system, models where convection is the dominant physical mechanism that sets the CE efficiency $\alpha_{\mathrm{CE}}$ are viable (however as discussed above, the predicted outcomes can vary dramatically depending on when during the AGB phase the CE phase begins). Future observations of other CE events with surveys such as the Vera C. Rubin Legacy Survey of Space and Time would be able to further elucidate the role of convection in setting CE efficiency \citep{2024arXiv240604118N}.
\end{enumerate}

Alessi12-PCE represents an important observational benchmark for linking pre- and post-CE stellar and orbital parameters, particularly at the high-mass end of WDs. However, a statistically meaningful understanding of the CE physics identified and examined in this work (e.g., the importance of timing of CE onset and the role of convection) requires a larger characterized sample of systems. At present, Alessi12-PCE is the only WD+MS post-CE binary in a star cluster where such constraints can be applied, making it difficult to robustly propagate these results to lower or higher mass systems. Expanding this sample to include systems spanning a wide range of WD masses, orbital separations, and evolutionary states is therefore a critical next step. 

Combined with targeted searches for post-CE binaries in star clusters using recent catalogs (e.g., \citetalias{Grondin2024}; \citealt{Li2025}; \citealt{2026ApJS..284...38Y}) and new observations, the framework presented in this work can be applied to a growing population of post-CE binaries in star clusters, enabling empirical constraints on CE physics that are inaccessible from field binaries alone.
Such a sample is currently in preparation, and will be essential for moving beyond heuristic prescriptions such as $\alpha_{\mathrm{CE}}$ and toward a more physically grounded understanding of CE evolution. Until then, we recommend that binary population synthesis efforts and hydrodynamical simulations discuss these key physical nuances of CE evolution when interpreting their results.

\section*{Acknowledgments}
The authors thank Marten van Kerkwijk, Kyle Rocha, Jay Strader, and Phil Van-Lane for their thoughtful feedback that improved this analysis and manuscript.
S.M.G. acknowledges the support of the Natural Sciences and Engineering Research Council of Canada
(NSERC), initially through an NSERC Postgraduate Scholarship—Doctoral (PGS-D) and currently though an NSERC Postdoctoral Fellowship (PDF). M.R.D. acknowledges support from NSERC through grant RGPIN-2025-06224, the Canada Research Chairs Program (CRC-2023-00127), the Ontario ERA program (ER22-17-164) and the Dunlap Institute at
the University of Toronto. J.N. and P.S.M. acknowledge support from U.S. National Science Foundation (NSF) award AST-2009713. J.N. also acknowledges support from U.S. NSF awards AST-2319326 and AST-2511139. P.S.M. also acknowledges support from the U.S. NSF award AST-2511138. A.L. acknowledges support from the NSERC and is funded through a NSERC Canada Graduate Scholarship—Doctoral. A.L. is also supported by the Data Sciences Institute at the University of Toronto through grant number DSIDSFY3R1P02. H.S. acknowledges partial salary support from a Moore Foundation Postdoctoral Fellowship Grant to Rutgers University.

The ZTF light curve in this work was obtained by \cite{https://doi.org/10.26131/irsa539}. This paper includes observations obtained at the international Gemini Observatory, a program of NSF NOIRLab, which is managed by the Association of Universities for Research in Astronomy (AURA) under a cooperative agreement with the U.S. National Science Foundation on behalf of the Gemini Observatory partnership: the U.S. National Science Foundation (United States), National Research Council (Canada), Agencia Nacional de Investigación y Desarrollo (Chile), Ministerio de Ciencia, Tecnología e Innovación (Argentina), Ministério da Ciência, Tecnologia, Inovações e Comunicações (Brazil), and Korea Astronomy and Space Science Institute (Republic of Korea). The Gemini data in this paper are from programs GN-2024A-Q-315 and GN-2024A-FT-109.

Some of the data presented herein were obtained at Keck Observatory, which is a private 501(c)3 non-profit organization operated as a scientific partnership among the California Institute of Technology, the University of California, and the National Aeronautics and Space Administration. The Observatory was made possible by the generous financial support of the W. M. Keck Foundation.

A major upgrade of the Kast spectrograph
on the Shane 3 m telescope at Lick Observatory, led
by Brad Holden, was made possible through generous
gifts from the Heising-Simons Foundation, William and
Marina Kast, and the University of California Observatories. Research at Lick Observatory is partially supported by a generous gift from Google.

\subsubsection*{Software Acknowledgment}

This work made use of the following software packages: \texttt{astropy} \citep{astropy:2013,astropy:2018,astropy:2022,astropy_17756022}, \texttt{Jupyter} \citep{2007CSE.....9c..21P,kluyver2016jupyter}, \texttt{matplotlib} \citep{Hunter:2007}, \texttt{numpy} \citep{numpy}, \texttt{pandas} \citep{mckinney-proc-scipy-2010,pandas_19340003}, \texttt{python} \citep{python}, \texttt{scipy} \citep{scipy_20764140}, \texttt{IRAF} \citep{1993ASPC...52..173T,1986SPIE..627..733T}, and \texttt{Lightkurve} \citep{2018ascl.soft12013L,Lightkurve_4654522}. The authors also acknowledge the use of Claude (Anthropic; Sonnet 5) to perform minor wording and data visualization edits.

This research has made use of the Astrophysics Data System, funded by NASA under Cooperative Agreement 80NSSC21M00561. This research also made use of the Canadian Advanced Network For Astronomy Research (CANFAR) operated in partnership by the Canadian Astronomy Data Centre and The Digital Research Alliance of Canada with support from the National Research Council of Canada the Canadian Space Agency, CANARIE and the Canadian Foundation for Innovation. This work uses Modules for Experiments in Stellar Astrophysics \citep[MESA;][]{Paxton2011, Paxton2013, Paxton2015, Paxton2018, Paxton2019, Jermyn2023}. Software citation information aggregated using \texttt{\href{https://www.tomwagg.com/software-citation-station/}{The Software Citation Station}} \citep{software-citation-station-paper}.

\bibliography{Alessi12, extra-bib}{}
\bibliographystyle{aasjournal}

\appendix
\renewcommand{\thefigure}{\thesection\arabic{figure}}
\renewcommand{\thetable}{\thesection\arabic{table}}
\setcounter{figure}{0}
\setcounter{table}{0}

\section{Summary of Observations}

\begin{figure*}[ht!]
    \centering
    \includegraphics[width=0.55\linewidth]{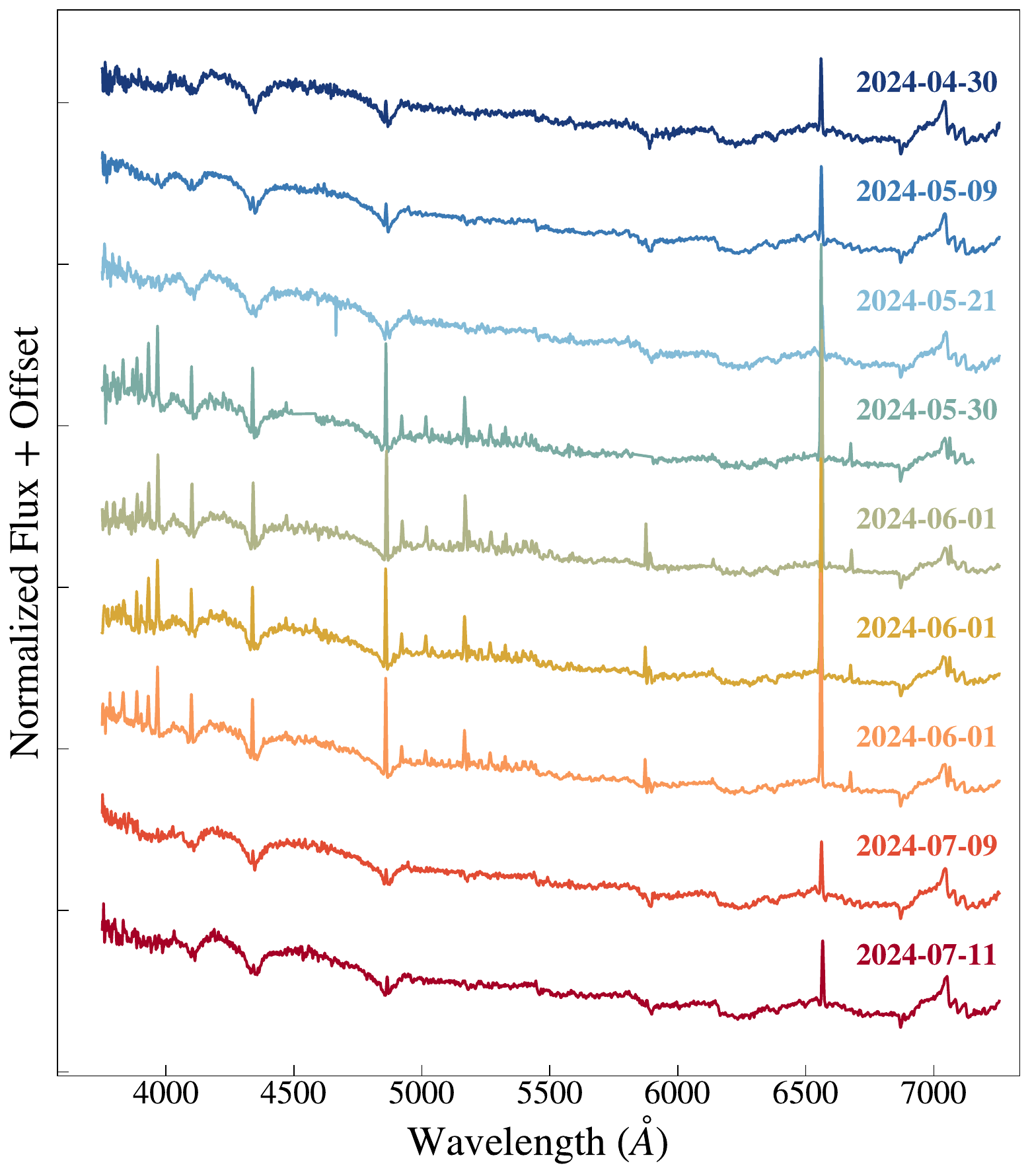}
    \caption{Gemini/GMOS spectroscopy of Alessi12-PCE. Each spectrum is constructed by stacking between two and four sequential exposures obtained during a single observing epoch (see Section~\ref{sec:geminispec}). We see an emission-line forest extending from $\approx$ 4900-5400\AA, which is present in a subset of the data. When comparing to \cite{Melis...2020ApJ...905...56M}, this appears to include Fe II lines at 2924 and 5018 \AA. There is another strong emission feature at $\approx$ 5870-5880\AA{}, which is consistent with He I D$_3$ emission, and could potentially be a signature of M dwarf activity \citep[][]{Medina...2020ApJ...905..107M}. Further analysis is required to conclusively characterize these features, but is beyond the scope of this work.} 
    \label{fig:gemini-allspec}
\end{figure*}

Here, we summarize the spectroscopic and photometric observations of Alessi12-PCE.
Table \ref{tab:obssummary} describes the spectroscopic observations obtained with
Shane/Kast, Gemini/GMOS, and Keck/LRIS. Table \ref{tab:uvot-obs} summarizes the
\textit{Swift}/UVOT photometric observations. Figure \ref{fig:gemini-allspec} shows
the multi-epoch Gemini/GMOS spectroscopy used to measure the WD and H$\alpha$ RVs.

\begin{deluxetable*}{lllll}[ht!]
    \tablecaption{Summary of spectroscopic observations of Alessi12-PCE. \label{tab:obssummary}}
    \tablewidth{0pt}
    \tabletypesize{\small}
    \tabcolsep=12pt
    \tablehead{
        \colhead{\textbf{Telescope}} & \colhead{\textbf{Instrument}} & \colhead{\textbf{Filter/Grating}} & \colhead{\textbf{Exposure Time (s)}} & \colhead{\textbf{Date}}
    }
    \startdata
    Lick/Shane & Kast & 600/4310 (blue), 600/7500 (red) & 5x1830 (blue), 10x900 (red) & 2023-10-24 \\
    Keck & LRIS & 600/4000 (blue), 400/8500 (red) & 1×900 (blue), 2x425 (red) & 2023-11-17 \\
    Lick/Shane & Kast & 600/4310 (blue), 600/7500 (red) & 1x1830 (blue), 2x900 (red) & 2023-12-05 \\
    Lick/Shane & Kast & 600/4310 (blue), 600/7500 (red) & 3x1830 (blue), 6x900 (red) & 2023-12-06 \\
    Gemini-N & GMOS & B480 & 4×705 & 2024-04-30 \\
    Gemini-N & GMOS & B480 & 4×690 & 2024-05-09 \\
    Gemini-N & GMOS & B480 & 4×690 & 2024-05-21 \\
    Gemini-N & GMOS & B480 & 2×690 & 2024-05-30 \\
    Gemini-N & GMOS & B480 & 3×690 & 2024-06-01 \\
    Gemini-N & GMOS & B480 & 4×690 & 2024-06-01 \\
    Gemini-N & GMOS & B480 & 4×705 & 2024-07-09 \\
    Gemini-N & GMOS & B480 & 4×705 & 2024-07-11 \\
    Gemini-N & GMOS & B480 & 4×705 & 2024-07-12 \\
    \enddata
\end{deluxetable*}

\begin{deluxetable*}{l|lcc|cc}[ht!]
    \tablecaption{\emph{Swift}/UVOT Photometry of Alessi12-PCE. \label{tab:uvot-obs}}
    \tablewidth{0pt}
    \tabletypesize{\small}
    \tabcolsep=3pt
    \tablehead{
        \colhead{Filter} & \colhead{Obs Date} & \colhead{Exp Time (s)} & \colhead{AB mag} & \colhead{Exp Time (s)} & \colhead{AB mag} \\
        \colhead{} & \multicolumn{3}{c}{Individual Exposures} & \multicolumn{2}{c}{Summed Image}
    }
    \startdata
    UVW2 & 2024-12-19T01:49:57 & 152.32 & 17.03 $\pm$ 0.09 & 627.47 & 16.95 $\pm$ 0.05\\
    UVW2 & 2024-12-19T22:13:44 & 475.15 & 16.93 $\pm$ 0.05 & \nodata & \nodata \\
    UVM2 & 2024-12-19T01:53:19 & 95.04 & 17.69 $\pm$ 0.13 & 388.49 & 17.68 $\pm$ 0.08\\
    UVM2 & 2024-12-19T22:23:57 & 293.45 & 17.68 $\pm$ 0.08 & \nodata & \nodata \\
    UVW1 & 2024-12-19T01:47:07 & 76.54 & 17.50 $\pm$ 0.13 & 314.50 & 17.44 $\pm$ 0.09\\
    UVW1 & 2024-12-19T22:05:25 & 237.96 & 17.43 $\pm$ 0.09 & \nodata & \nodata \\
    U & 2024-12-19T01:48:29 & 38.16 & 18.17 $\pm$ 0.14 & 157.02 & 18.35 $\pm$ 0.09\\
    U & 2024-12-19T22:09:32 & 118.86 & 18.43 $\pm$ 0.10 & \nodata & \nodata \\
    B & 2024-12-19T01:49:13 & 38.16 & 18.52 $\pm$ 0.29 & 157.03 & 18.54 $\pm$ 0.14\\
    B & 2024-12-19T22:11:37 & 118.87 & 18.54 $\pm$ 0.16 & \nodata & \nodata \\
    V & 2024-12-19T01:52:36 & 38.17 & $>$ 18.05 & 157.04 & $>$18.91 \\
    V & 2024-12-19T22:21:52 & 118.87 & $>$ 18.75 & \nodata & \nodata \\
    \enddata
\end{deluxetable*}

\end{document}